\documentclass[information,article,accept,pdftex,moreauthors]{MDPIDefinitions/mdpi} 
\firstpage{1} 
\pubvolume{1}
\issuenum{1}
\articlenumber{0}
\pubyear{2022}
\copyrightyear{2022}
\externaleditor{Academic Editor: {Willy Susilo} 
}
\datereceived{27 June 2022} 
\dateaccepted{28 July 2022 } 
\datepublished{} 
\hreflink{https://doi.org/} 

\usepackage{framed}

\usepackage{amssymb}
\usepackage{latexsym}

\definecolor{newcolor}{rgb}{.8,.349,.1}

\usepackage{syntax}

\usepackage[labelformat=simple]{subfig}

\newcommand{\squishlist}{
	\begin{list}{$\bullet$}
		{ \setlength{\itemsep}{0pt}      \setlength{\parsep}{3pt}
			\setlength{\topsep}{3pt}       \setlength{\partopsep}{0pt}
			\setlength{\leftmargin}{1.5em} \setlength{\labelwidth}{1em}
			\setlength{\labelsep}{0.5em} } }
	
	\newcommand{\squishlisttwo}{
		\begin{list}{$\bullet$}
			{ \setlength{\itemsep}{0pt}    \setlength{\parsep}{0pt}
				\setlength{\topsep}{0pt}     \setlength{\partopsep}{0pt}
				\setlength{\leftmargin}{2em} \setlength{\labelwidth}{1.5em}
				\setlength{\labelsep}{0.5em} } }
		
		\newcommand{\squishend}{
		\end{list}  }

\Title{\textls[-20]{Q4EDA: A Novel Strategy for Textual Information Retrieval Based on User Interactions with Visual Representations of Time Series}}

\TitleCitation{Q4EDA: A Novel Strategy for Textual Information Retrieval Based on User Interactions with Visual Representations of Time Series}

\Author{Leonardo Christino 
 \orcidA{}, Martha D. Ferreira \orcidB{} 
 and Fernando V. Paulovich *\orcidD{}}

\AuthorNames{Leonardo Christino, Martha D. Ferreira, and Fernando V. Paulovich}

\AuthorCitation{Christino, L.; Ferreira, M.; Paulovich, F.;}

\address[1]{%
Faculty of Computer Science, Dalhousie University, 
Halifax, NS B3H 1W5, 
 Canada; christinoleo@dal.ca (L.C.); dais.martha@dal.ca (M.D.F.)}

\corres{\hangafter=1 \hangindent=1.05em \hspace{-0.82em} Correspondence: paulovich@dal.ca; Tel.: +31-6-27479789}

\abstract{Knowing how to construct text-based \textit{Search Queries (SQs)}
 for use in \textit{Search Engines (SEs)} such as Google or Wikipedia has become a fundamental skill. Though much data are available through such SEs, most structured datasets live outside their scope. Visualization tools aid in this limitation, but no such tools come close to the sheer amount of information available through general-purpose SEs. To fill this gap, this paper presents Q4EDA, a novel framework that converts users' visual selection queries executed on top of time series visual representations, providing valid and stable SQs to be used in general-purpose SEs and suggestions of related information. The usefulness of Q4EDA is presented and validated by users through an application linking a Gapminder's line-chart replica with a SE populated with Wikipedia documents, showing how Q4EDA supports and enhances exploratory analysis of United Nations world indicators. Despite some limitations, Q4EDA is unique in its proposal and represents a real advance towards providing solutions for querying textual information based on user interactions with visual representations.}

\keyword{\textls[-25]{information retrieval; visual analytics; search engine;
 \emph{visual selection query}; visual information} retrieval; exploratory analysis 
} 

\begin{document}
\section{Introduction}
\label{sec:intro}

It is undeniable that advancements in \textit{{Search~Engines~(SE)}}~\cite{croft2010search} represent a revolution without precedent in human history for information dissemination. Due to the large amount of readily available data, knowing how to construct \textit{{Search~Queries~(SQs)}}~\cite{sqsdef} has become a fundamental skill. From Google~\cite{google} to Wikipedia~\cite{wikipedia}, most SEs require users to type a text-based search query to retrieve relevant information. Although much data are available through such SEs, most of the structured datasets, such as the UNData~\cite{undata} time-series world indicators, live outside their scope and cannot be incorporated into search queries. On the other hand, the information contained in structured datasets is typically available to probe through visualization tools or interfaces. One example is Gapminder~\cite{rosling2012gapminderorg}, where users can visualize animated charts displaying the UNData dataset. Despite their popularity, no individual visualization tool comes close to the sheer amount of information available through Wikipedia's hyperlinked text documents. Therefore, even if one uses, for instance, Gapminder to discover global misconceptions~\cite{sarma2017hans}, they will undeniably be required to search elsewhere for extra information regarding the underlying findings.

The task of generating text-based SQs for SEs is, broadly speaking, performed by the users themselves. However, this process can be challenging since the translation between user intent and keywords is sometimes not trivial, causing users not to find the information being looked for~\cite{kammerer2012children}. Beyond regular keyword lists, other search strategies are available to address such limitations. For instance, it is also possible to ``search by image''~\cite{reilly2017reverse}, where users can find images similar to a given image, or ``search by natural language''~\cite{cafarella2005search, kammerer2012children}, where users can search using descriptive texts. 

However, when considering structured datasets, up to our knowledge, there is no mechanism to use pieces of visual data, such as user selections in a line chart, to search for related information using general-purpose SEs. Instead of using visual selections, there are solutions that either use visualizations as outputs~\cite{hullman2013contextifier, badam2018elastic, yu2019flowsense} or as a visual interface to manually construct what is essentially a text-based query~\cite{kraska2018northstar, zhou2021modeling, borland2019selection, borland2020selection}. Even though such visualizations are shown to aid the information gathering processes, in these works, the use of visual selections as \textit{{visual search queries}} to fetch relevant information from general-purpose SEs is an unexplored concept. Various examples of search query suggestions for data analysis~\cite{ooi2015survey, yi2017autog} have also been proposed, but the use of such techniques applied to \textit{{visual search queries}} during data analysis is similarly unexplored.

This paper proposes \textit{{QuEry for visual Data Analysis (Q4EDA)}}, a novel framework that converts user visual selections to relevant search queries to be used for general purposes search engines, such as Google or Wikipedia. 
Using query expansion and suggestion techniques, one of the promising applications for Q4EDA is to allow users to perform an enhanced visual analysis of time-series dataset collections. To use Q4EDA in this context, a visualization tool, such as Gapminder or Tableau~\cite{zhang2012visual} captures and forwards user \textit{visual selection queries (VQs)} to Q4EDA, then Q4EDA processes the selection and outputs an SQ which can then be used on general-purpose SEs. Q4EDA also suggests other potential time series related to the selected event within the dataset collection to be subsequently investigated. In summary, the main contributions of this paper are:

\begin{itemize}[leftmargin=*,labelsep=5.8mm]
\item A conversion technique to transform visual selection queries into valid and stable search queries usable in general-purpose search engines; 

\item \textls[-15]{A strategy to expand the converted search query to better retrieve related text documents and provide suggestion ranking lists with data related to the visual selection~query;}

\item An exploratory data analysis application example that uses Q4EDA in practice to provide means to find more (textual) information related to an observed pattern compared to the standard manual keyword-based queries.
\end{itemize}

%
%

%

The remainder of the paper is structured as follows. In Section~\ref{sec:related}, we discuss related work involving techniques that seek to interpret visual interactivity as queries and how they execute and use the results of their query. In Section~\ref{sec:method}, we formalize the problem and outline the Q4EDA solution. In Section~\ref{sec:results}, we present use-case examples of how visualization tools can use Q4EDA. Evaluations through user survey and query stability performance are then presented in Section~\ref{sec:usereval}, indicating a good degree of stability and reproducibility of the search queries and overall success of Q4EDA in enhancing users' ability to probe for relevant information of patterns or events found during visual data analysis. Finally, in Section~\ref{sec:limitations} we discuss Q4EDA limitations and in Section~\ref{sec:conclusions} we draw our~conclusions.

\section{Related Work}
\label{sec:related}

In order to support users in visual data analysis, some tools and techniques allow for a non-obstructive exploratory approach through visual interactivity~\cite{srinivasan2018augmenting, kraska2018northstar}, among the most relevant is NorthStar~\cite{kraska2018northstar}, which goes in-depth into the difficulties of providing an exploratory system that follows responsive and real-time guidelines for intuitive and engaging user exploratory analysis while at the same time utilizing automatic problem detectors throughout the entire workflow to reduce the amount of potential bias or incorrect insights generated through the exploration. However, although a certain level of ``query conversion'' is performed through these tools or techniques, they still expect what is essentially a text-based query to be constructed manually through their visual interface. Furthermore, these tools use the query to retrieve data from domain-specific databases, which differs from our proposed query conversion method for search engines. Moreover, although research has identified other ways to encode visual findings and hypotheses beyond visual query constructors~\cite{suh2022grammar}, no existing work, as far as the authors know, has demonstrated ways to use such methodology to convert a visual selection into a search query as Q4EDA~proposes.

Considering the perspective of visual selection queries and its related information from external datasets, some approaches focus on extracting or generating automatic visualizations and annotations. Despite representing significant progress, the amount of analysis enhancement they provide is limited to the generation of visual metaphors~\mbox{\cite{cui2019text,lin2018vizbywiki,bryan2016temporal}} and of visual feature extractions~\cite{bryan2016temporal,tang2017extracting,ding2019quickinsights} produced from a single dataset used for analysis. Other works focus on generating enhanced visualizations by displaying one dataset as a visual feature of the other. These visual features can be annotations~\cite{hullman2013contextifier,kwon2014visjockey}, can involve textual queries to generate visualizations~\cite{yu2019flowsense,luo2018deepeye,metoyer2018coupling, hoque2017applying}, can define a question-answer interface to visualizations~\cite{kim2020answering, kafle2020answering, yu2020cross}, can use textual datasets to help the understandability of structured datasets~\cite{badam2018elastic,kim2018facilitating,yu2019flowsense,srinivasan2017orko}, or can automatic link text to images for use within visualizations~\cite{mogadala2019trends,yu2020reasoning,dhelim2020compath}. The research has shown approachable and engaging ways to optimize visualizations and analysis by using text-based datasets and NLP techniques. Of them, usage of heterogeneous data~\cite{dhelim2020compath} an cross-modal methods~\cite{yu2020reasoning,yu2020cross} has shown that utilizing multimodal datasets, such as time-series and text, provides significant advantages to the user's analysis. Although many of them use some query to retrieve information from text, no one uses external search engines as the target of such queries. Instead, they focus on specific text datasets. These efforts show a focus on analyzing a single dataset with the aid of another, which provides NLP capabilities during analysis. However, such approaches differ from the proposal of Q4EDA where we do not just convert visual search queries and provide valid search queries to be used in existing search engines but also provide query suggestions, aiding the analysis in many ways. 


Visual Analytics has also recently started to use heterogeneous datasets for analysis, and one such work performed by~\cite{zhou2021modeling} proposes a method of using users' interactions to quantify their attention and, from it, decide which medical documents to present to the user. Arguably similar to Q4EDA, \cite{zhou2021modeling} uses a textual database and a query system from their previous work Cadence~\cite{borland2019selection,borland2020selection}. Although their text database does not significantly differ from existing search engines, hence approaching Q4EDA's proposal, this work does not provide an actual query conversion process, nor propose methods of query suggestions, nor allow for visual selections as queries. Instead, queries are again constructed manually using the Cadence system through a drag-and-drop interface to perform the search. This procedure is arguably most similar to the interface of NorthStar~\cite{kraska2018northstar} as opposed to Q4EDA.

Significant effort has been made by the information retrieval community to propose better ways to use search engines during data analysis. The practice of retrieving information~\cite{croft2010search} describes ways to expand a query, so its effectiveness is more relevant to users when executed in a SE~\cite{zhang2009concept,carpineto2012survey,azad2019query,dahir2021query} and how to provide suggestions for future queries~\cite{ooi2015survey}. Although their techniques are very relevant to Q4EDA, the vast majority are non-visual approaches that describe advances in how to perform query expansion of a keyword or natural language search query as opposed to Q4EDA's \textit{{visual selection queries}}. Among the ones that use visualizations, they either use it only to display information, lacking interactivity~\cite{hoeber2005visualization}, or use it as a query builder interface, where there is no concept of visual selection queries~\cite{khazaei2017supporting}, focuses on visualizing a manually or interactively constructed query~\cite{scells2018searchrefiner,russell20182dsearch}. Indeed, the authors are not aware of any research within the information retrieval community which transforms or converts visual selections into SQ automatically, that is, without any user intervention to manually, even if through visual interfaces, construct the search query. Q4EDA aims to fill this gap where users are not required to interpret the available data and manually construct the search query. Instead, users can visually select portions of existing visualizations, and the selection is automatically converted into a valid search query for use in search engines.

Both information retrieval and data management communities have had great strides in connecting heterogeneous data, such as time-series datasets and text documents. Currently, one of the major advances in this direction is the concept of dataspaces~\cite{curry2020dataspaces,franklin2005databases} and knowledge graphs~\cite{dahir2021query,balalau2020graph,martinez2015automated,auer2007dbpedia,yi2017autog}. Manual, semi-automatic, and automatic procedures for matching and linking these have been studied~\cite{golshan2017data,mountantonakis2019large,mogadala2019trends,arya2021survey,christophides2019end,groger2014deep, roy2005towards}. 
Though some contributions links structured datasets within visualizations to external data and others link data to text documents, no related work performs the full path from visual selections to search engine responses as far as the authors are aware. Though Q4EDA uses many of the concepts shared among these valuable contributions, Q4EDA blurs the line between visual interaction, database query, and information retrieval to deliver what it proposes.

In summary, existing works provide many relevant works which use textual datasets to enhance data analysis, but they provide no \textit{{visual selection query}} conversion capabilities. Many do not support search engines nor provide query suggestions from text documents extracted from said search engine. Some newer work exemplifies the benefits of linking structured datasets to textual counterparts for data analysis enhancement. However, they still fail to provide a way for users to analyze datasets through visual selection queries, such as a box selection within a line chart, and simultaneously be provided suggestions of said visual selection. Instead, existing work either fails to provide the visual selection query, the query conversion, the support for SEs, or the query suggestions. This is the novelty of our framework: to provide a visual analytic workspace where external VA tools provide enhanced data analysis of time-series datasets by converting visual selection queries to search queries to be used within well-known SEs and supplying query suggestions to further enhance the analysis. To promote this query conversion, we devise a novel approach using existing pattern analysis and natural language process strategies to convert user-selected findings and elements of interest to valid SQs while also providing suggestions based on the selected finding, allowing users to navigate the data and build up knowledge.

\section{Q4EDA}
\label{sec:method}

In this paper, we present \textit{{QuEry for visual Data Analysis (Q4EDA)}}, a novel query conversion framework designed to enhance data exploration tasks by ingesting user's findings through visual selections and returning \textit{{search queries (SQs)}} which can retrieve relevant information from general-purpose \textit{{search engines (SEs)}}. By identifying user-driven visual selections of findings (or patterns) of interest, Q4EDA allows visualization tools to request the conversion of said selections into SQs to be used in keyword-based SEs to retrieve textual information related to the finding. Additionally, Q4EDA provides query suggestions with recommendations based on the similarity or correlation of the selected finding to other parts of the dataset under analysis. Q4EDA is specifically designed for time-series dataset collections, hence it expects the visualization tools to similarly include time-series data as the input visual selection query.

\subsection{Overview}\label{sec:overview}

\begin{figure}[H]

\begin{adjustwidth}{-\extralength}{0cm}
\centering 
    \subfloat[\centering] {\includegraphics[width=0.6\textwidth]{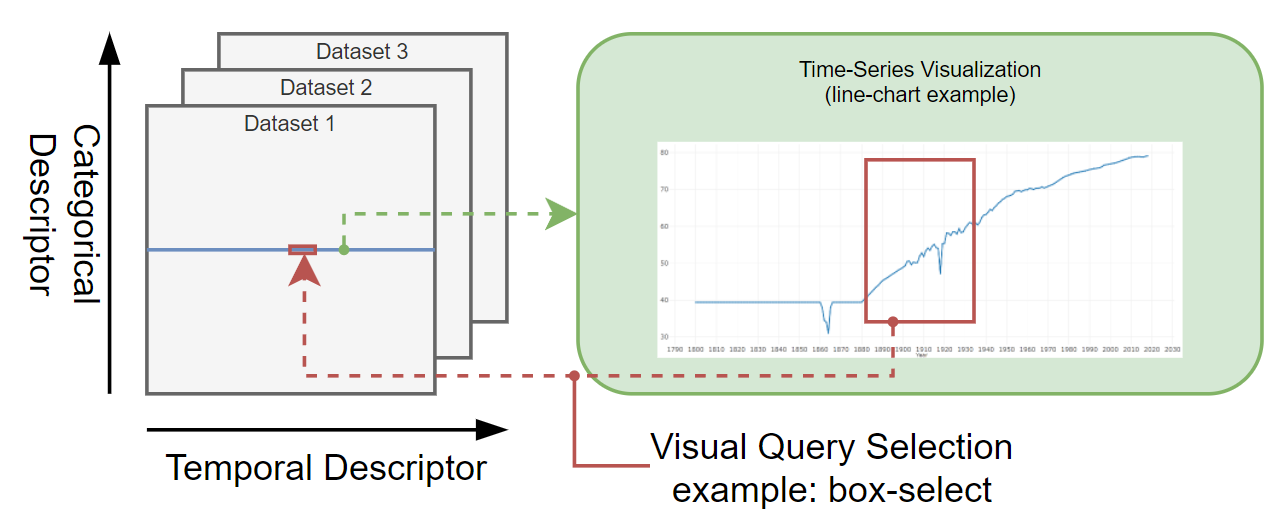}\label{fig:vqoverview}} 
    \subfloat[\centering] {\includegraphics[width=0.6\textwidth]{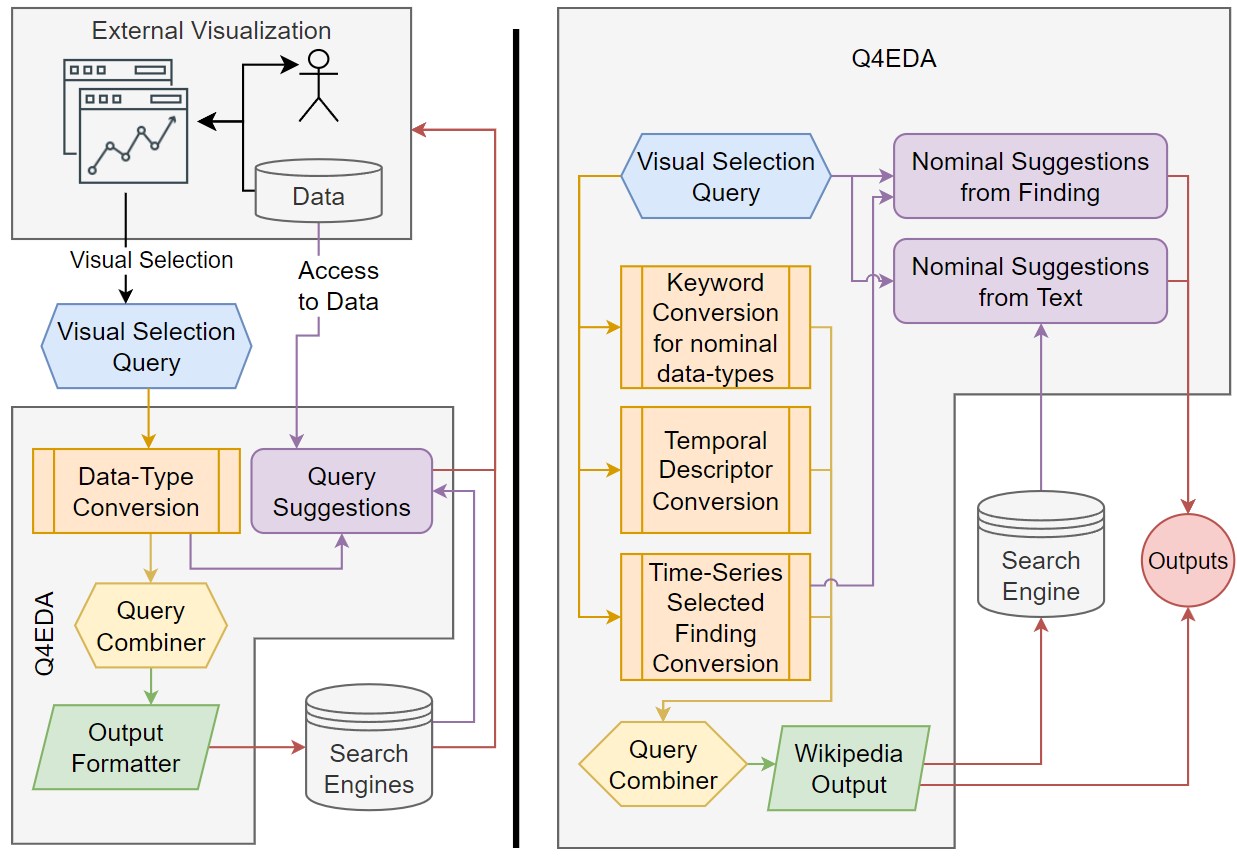}\label{fig:q4edaoverview}}
\end{adjustwidth}
    \caption{Overall summary of Q4EDA. (\textbf{a}) Time-series dataset collection structure expected by Q4EDA (left) is composed of multiple time-series datasets. Each row of the many datasets are the time-series which are visualized through tools such as Gapminder~\cite{gapminderusaleviz}, and a selection within such visualization (in red) comprises the \textit{{Visual Selection Query (VQ)}} converted by Q4EDA. (\textbf{b}) Overview of Q4EDA conversion and suggestion processes. Q4EDA expects to have access to a dataset collection and a given user's VQ (left), and the inner operations convert each data-type into output queries which are combined as a valid SQ (right). After executing the SQ in a SE and retrieving the text documents, Q4EDA also provides query suggestions.}
    \label{fig:overview}
\end{figure}

Q4EDA is structured as summarized in Figure~\ref{fig:overview}. After a setup phase, a visualization may forward a \textit{{Visual Selection Query (VQ)}} to Q4EDA to be processed and converted into a \textit{Search Query (SQ)}. Q4EDA can be divided into three distinct steps: conversion, output, and suggestion. Q4EDA conversion uses the VQ's data-types, 
 such as numeric or categorical values. Similarly, each of Q4EDA's outputs implements the SQ specification of a target SE. Finally, Q4EDA provides query suggestions by either correlating text documents retrieved from the SE to the dataset's available data or correlating the selected time-series pattern to other available time-series within the dataset. Although what we present of Q4EDA's implementation attempts to emulate the data available within Gapminder, the framework is purposely flexible to allow for other applications, such as will be discussed.

\subsection{Design Requirements}
\label{sec:design}

Q4EDA focuses on providing a VQ conversion framework for time-series dataset exploratory analysis. By combining elements of information retrieval, the gaps in the literature in regards to VQ conversion, and the goal stated above, we have compiled a list of requirements:

\begin{description}
\item[R1---Visual Query Conversion.] 
 Q4EDA must provide a valid SQ given a VQ that provides: (a) \textbf{output correctness}---the information returned from the targeted SEs should be related to the user's VQ~\cite{croft2010search}; and (b) \textbf{output stability}---the results should not vary too much given slight variations on the VQ since slight variations of user interaction (selection) are expected to happen in practice~\cite{carpineto2012survey};
%
%

\item[R2---Information Retrieval.] Q4EDA should follow existing approaches to enhance the retrieved results, including: (a) \textbf{query expansion}---the converted SQ should be expanded~\cite{azad2019query} with terms to broaden the query results to better allow for related information to be found; and (b) \textbf{query suggestion}---suggestions~\cite{ooi2015survey} should be provided to provide users with information of interest.
%
\end{description}

\subsection{Q4EDA Definitions}
\label{sec:inputformal}

As a first step, it is essential to define what an SQ is in our context. Many online resources define SQ as ``A search query or search term is the actual word or string of words that a search engine user types into the search box''~\cite{keyvssq}. Beyond that, we see a plethora of similar concepts when applied to different modes of communication. To simplify and better contextualize Q4EDA, we define a \textit{Search Query (SQ)} as a term-based format of querying for information. Similarly, we define a \textit{Search Engine (SE)} as any system which can receive an SQ and retrieve information relevant or related to the query from its database(s), where among the most famous examples are Google, Wikipedia, Apache Lucene and Elasticsearch~\cite{elasticsearchwikipedia}.


As we discussed, Q4EDA is designed to be used by a data analysis visualization tool to convert a \textit{visual selection query (VQ)} into a \textit{search query (SQ)}, which can then be used to retrieve information from a \textit{search engine (SE)}. For this, Q4EDA expects to have access to the time-series dataset collection 
being analyzed. Q4EDA also expects the data to be organized as such: the dataset collection is made up of multiple datasets wherein lives the many time-series, as is exemplified in Figure~\ref{fig:overview}a. For instance, UNData~\cite{undata} is a dataset collection where multiple datasets of so-called indicators (e.g., ``life expectancy'' or ``child mortality'') contains a per country time-series. With this, we have the four main \textit{data-types} of Q4EDA conversion: \textit{dataset name}, \textit{categorical descriptor} {such as country name}, \textit{temporal descriptor} and \textit{time-series value}. With this, we also define a \textit{Visual Selection Query (VQ)} as a subset of a visualization's data, which was selected through user interaction as exemplified as the red selection of Figure~\ref{fig:overview}a. For instance, Gapminder's line-chart~\cite{gapminderusaleviz} could theoretically allow for a box-selection or lasso-selection of a part of the displayed data with which the user can indicate a visual selection query. This VQ would include the selection years as the x-axis's temporal descriptor of the selection, the y-axis's time-series numeric values and the corresponding dataset name and categorical descriptor of that individual line-chart.

With these definitions in hand, we can describe Q4EDA as a framework that receives a VQ and, by utilizing the available dataset, it converts the VQ into one of the output SQ formats. After executing said SQ, Q4EDA also provides query suggestions among the available dataset names and categorical descriptors. For that, the workflow to use Q4EDA follows the representation of Figure~\ref{fig:overview} where after a user performs a VQ, its data are given to Q4EDA for the query conversion and suggestion process.

\subsection{Query Conversion}
\label{sec:queryspec}

Q4EDA's framework leans on the three aspects: \textit{query conversion}, \textit{query combiner}, and \textit{query suggestion}, as is seen in Figure~\ref{fig:overview}. While the query combiner defines the supported SEs, the query conversions define the process used to convert each of VQ's individual data-types to generate a relevant output SQ using query expansion techniques (\textit{R2.a}). For this, Q4EDA first converts each data-type separately, and then the results are \textit{combined} and \textit{formatted} into the output SQ.

The inner mechanism used by Q4EDA to process and store the individual conversion step follows a grammatical convention based on existing SQ grammars, the closest of which is Elasticsearch's Simple Query format~\cite{SimpleElastic}. The formal grammar described in EBNF~\cite{feynman2016ebnf} is:

\setlength{\grammarparsep}{1pt plus 1pt minus 1pt} 
\begin{grammar}\label{grammar}
<sub-expression> ::= <or> | <and> | <required> | <term>

<or> ::= `(' <sub-expression> \{ `|' <sub-expression> \}+ ')'

<and> ::= `(' <sub-expression> \{ `&' <sub-expression> \}+ ')'

<required> ::= `\"'  <sub-expression> `\"'

<term> ::= [<negative-factor>] <inner-term> [<weight-factor>]

<inner-term> ::= `(' <spaced-term> `)' | <word>

<spaced-term> ::= <spaced-term> ` ' <word> | <word>

<word> ::= \{ <lower-case-letter> \}+

<weight-factor> ::= superscript ? <weight-factor>

<negative-factor> ::= `-',
\end{grammar}
where 
 the conversion's \textit{sub-expression} output consists of multiple \textit{terms} which can represent the input \textit{positively} or \textit{negatively}, and with the added \textit{weights} and logical operations, the output is able to be descriptive enough for formatting to target many SEs. In order to exemplify the grammar, if we consider an input which is requesting Q4EDA to convert the question ``population of the United States'' into a SQ, a plausible output could be $\{ (\text{united states} \mid \text{usa} \mid \text{america} \mid (\text{north america})^{0.5}) \& (\text{population} \mid \text{habitants} \mid \text{people}^{0.5} \mid (-\text{death})^{0.5} ) \}$, where it includes perfect match terms (e.g., ``united states''), terms with lower weights for less exact matches (e.g., ``North America'') and negative terms to indicate opposite meaning or antonyms (e.g., ``death'').

\subsubsection{Keyword Conversion}
\label{sec:indicatortags}

The first conversion tackles the keyword data-type, such as the dataset name or categorical descriptors. By acquiring the dataset name(s) from a given VQ or the categorical descriptor(s), Q4EDA uses natural language processing to generate a set of related terms for inclusion in its output. We first apply a text mining approach to assign a set of related terms $T^D_d=\{t_1^d, t_2^d, \ldots\}$ for every keyword $d$ received. For instance, the keyword $d=\text{``life expectancy''}$, which is a dataset name from UNData~\cite{undata}, besides of being represented by the terms ``life'' and ``expectancy'', can also be represented by terms such as ``longevity'' and ``lifetime''. Similarly, the same keyword is negatively represented by terms such as ``mortality'' or ``death''. Such an example results in the terms $T^D_{life\_expectancy}=\{life, expectancy, longevity, lifetime, -mortality, -death\}$ where positive terms are related to the keyword, and negative terms are negatively related, a concept similar to antonyms. In this way, any such nominal descriptors will have their semantic meaning defined by their associated terms. Finally, the set of terms are formatted into an expression following Q4EDA's grammar: $(life \| expectancy \| longevity \| lifetime \| -mortality \| -death)$.

This operation is performed through a pre-trained GloVe model~\cite{pennington2014glove} which generates related tags for a given keyword. First, the GloVe model is loaded using the Gensim library~\cite{rehurek2011gensim}, and then every unique keyword $d$ is tokenized and lemmatized with NLTK~\cite{loper2002nltk} and WordNet~\cite{fellbaum2010wordnet, zhang2009concept} and passed to Gensim's ``similar\_by\_vector'' method.

\subsubsection{Country Keyword Conversion}
\label{sec:geolocal}

Although the simple method described so far works well for dataset names and categorical descriptors with common nouns, specialized versions of keyword conversion are used to dataset names or categorical descriptors with proper nouns, such as geo-locations such as street, building name, city, state or country and events such as ``the second world war'', for better query expansion results~\cite{bhogal2007review}. Therefore, Q4EDA provides one of such specialized keyword conversions to properly process a country keyword.

Since geo-information also encodes geographical concepts such as continents, distances, area, and population, among others, this specialized conversion can output a more relevant set of terms for a given country. That said, this module design could expand indefinitely due to its data complexity, therefore we limited our processing to grammatical variations, naming conventions, synonyms, and regionality. To create the set of terms $T^C_c$ of words for a country keyword, the country's name is used along with other terms related to the country, such as adjectives or nouns (e.g., ``United States'' adds ``America'', ``American'' and ``USA''). Although a VQ may be specific for a single country, the VQ's relevant information within the target SE may include other similar or neighboring countries as well. Therefore, we add terms for the collection of countries nearby the selected country, such as the continent name or sub-area, with a smaller weight (e.g., ``United States'' adds ``North America'' with lower weights). This generalization has shown to be essential because some textual information may only contain more general references to the geographical location where a specific event occurred. To complement $T^C_c$ with said extra data, we extracted data from Gapminder's geography dataset~\cite{gapmindergeography} and used it as a reference for any country \textit{categorical descriptor}, such as is the case with UNData. Therefore, such an example results in the following output:
\begin{align*}
T^C_{united\_states} = &(\text{united states} \mid \text{united states of america} \mid \\
& \text{american} \mid \text{america} \mid \text{usa} \mid (\text{north america})^{0.5}).
\end{align*}

Finally,  
the list of terms is converted to Q4EDA inner grammar definition.

\subsubsection{Temporal Descriptor Conversion}
\label{sec:temporal}

Unlike previous processors, the temporal descriptor conversion is unique since it tackles a continuous temporal value and has some inner context that can be used to enhance search queries. Similar to the country keyword conversion, this step is required to be specialized to a given temporal variable, such as date, time, month, week-day, year or any combination of them. Therefore, Q4EDA provides one such implementation which focuses on converting a year range into a valid query output to be included in Q4EDA's output SQ.

The term set generated has the form of $T^E_{y_a, y_b}=\{t_1^e, t_2^e, \ldots\}$ where $y_a$ and $y_b$ identify the limits of the range of years. The natural terms associated with a range of years are the years themselves. However, this conversion process goes beyond and allows the VA Tool to define a weight distribution to better describe the selection interest within the range of years. For instance, while the default weight profile will give the same weight to all years within a range, if the VA Tool includes a gaussian weight distribution, the processor will give higher weights to the center of the range and lower to the others.

Additionally, the year conversion compares the year range to a predefined year range term. For example, if the year range is 1950--1960, an additional term of 1950s is included to represent the decade selected. If the year range does not match the exact decade, lower weights are given to the decade terms. The weight used in decade terms is $w=1-2\times(|ds^b_a| + |ds^e_b|)$ where 
 $ds^b_a$ is the distance from the first year to the beginning of the decade and $ds^e_b$ is the distance in years from the last year to the end of the decade, and weights equal or below $0$ causes the decade term to be discarded. The same process is applied for centuries. Therefore, such an example results in the following output:
\begin{align*}
T^E_{1851--1859, gaussian} =& (1851^{0.2} \mid 1852^{0.3} \mid 1853^{0.5} \mid 1854^{0.8} \mid 1855 \mid  \\
& 1856^{0.8} \mid 1857^{0.5} \mid 1858^{0.3} \mid 1859^{0.2} \mid 1850s^{0.6}).
\end{align*}

\subsubsection{Time-Series' Selected Finding Conversion}
\label{sec:patterndetection}

\begin{figure}[H]

\begin{adjustwidth}{-\extralength}{0cm}
\centering 
    \subfloat[\centering] {\includegraphics[width=8cm]{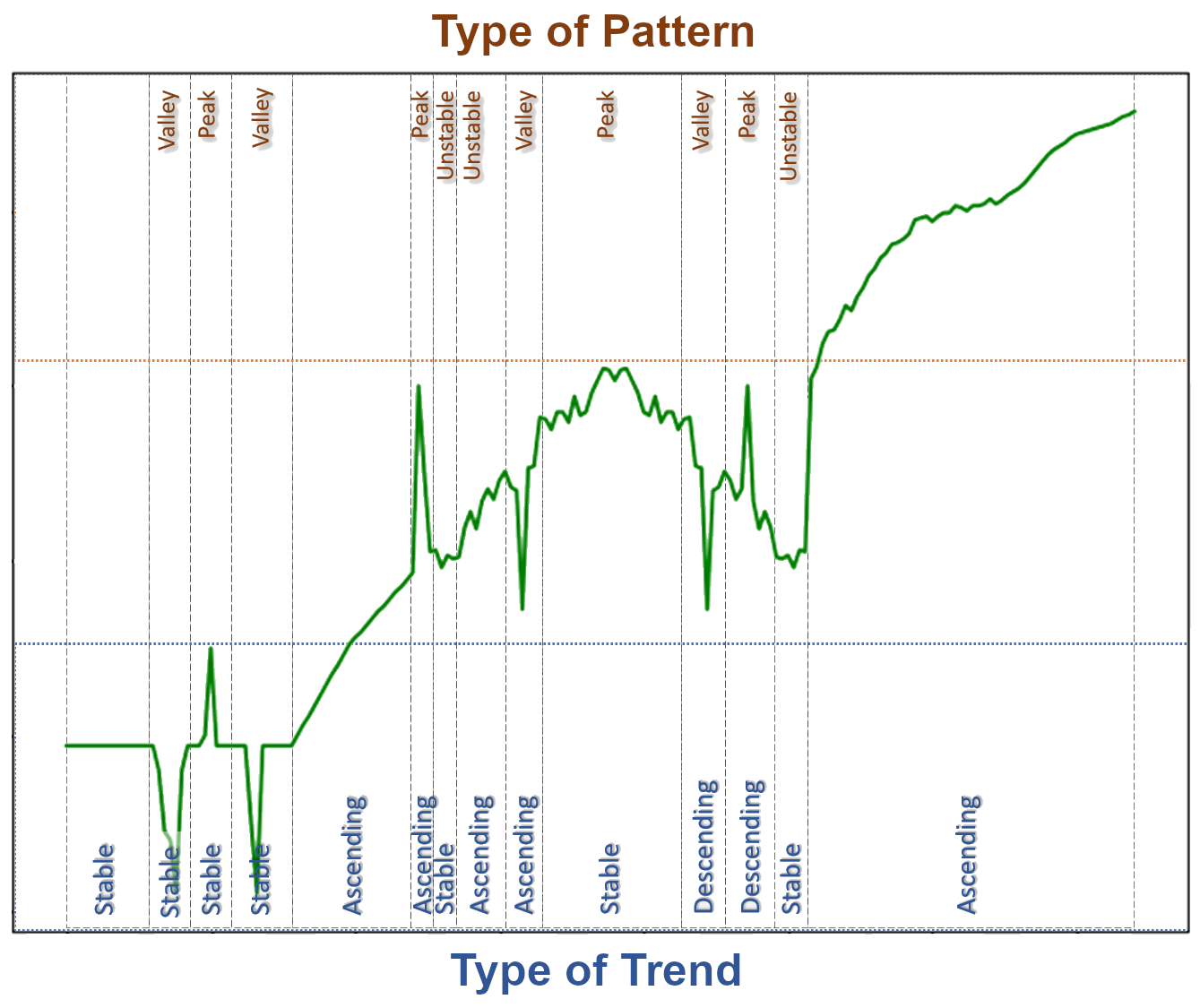}\label{fig:detectionpossibilities}}\quad
    \subfloat[\centering] {\includegraphics[width=8cm, height=6.8cm]{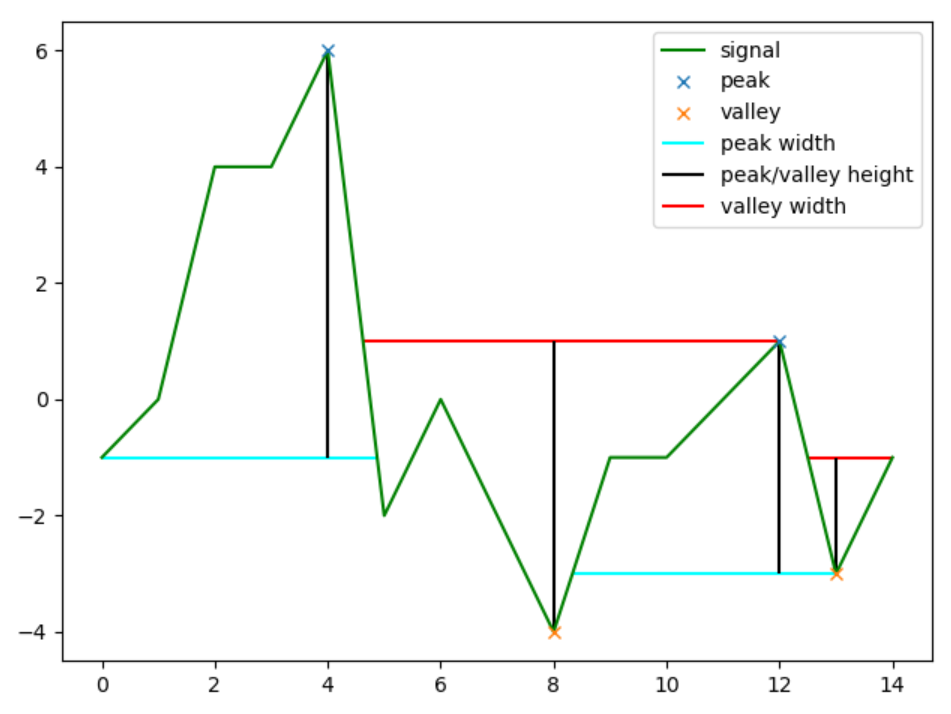}\label{fig:patternpossibilities}} 
\end{adjustwidth}
    \caption{\textls[-15]{Examples of trends and patterns combinations and peaks and valleys detected by our algorithm. (\textbf{a}) Combinations of trends and pattern types in visual selections. (\textbf{b}) Peaks and valleys~detection.}}
    \label{fig:patternpossibilitiesoverview}
\end{figure}

Finally, we analyse the numerical values selected by the user $F=I^d_c(y_a, y_b)=\{e_{y_a}, \ldots, e_{y_b}\}$ and converts the underlying pattern into a set of terms $\{t_1^p, t_2^p, \ldots\}$. Our implementation categorizes the selected values, which we call \textit{finding}, by its trend $tr$, which can be either \textit{ascending}, \textit{descending} or \textit{stable}, and by its pattern $tp$, which can either be \textit{peak}, \textit{valley}, and \textit{neutral}. This results in nine different possible trend/pattern combinations, as exemplified by Figure~\ref{fig:patternpossibilitiesoverview}a. These two identifiers are applied to a finding by using existing trend/pattern statistical methods and the resulting keyword is then converted to the desired query output using the same GloVe model~\cite{pennington2014glove} NLP process used in the Keyword Conversion (see Section~\ref{sec:indicatortags}). That said, if the input has only one value (e.g., $a = b$) or 
 no value, this conversion step will output nothing at all.

\vspace{-12pt}

\textls[-5]{The type of a selected trend $tr$ is defined by applying the Moving Averages (MA) method, a well-known technique in time-series analysis to define whether a series is stationary or not, providing us with a trend estimation~\cite{brockwell2016introduction}. MA first transforms the selection $F=I^d_c(y_a, y_b)=\{e_{y_a}, \ldots, e_{y_b}\}$ into a new series $F'=I'^d_c(y_a, y_b)=\{e'_{y_a}, \ldots, e'_{y_b}\}$ setting its values $e' \in I'^d_c(y_a, y_b)$ to the average value in the time interval $e'_p(w) = \mu( e_{y-w}, \ldots, e_y, \ldots, e_{y+w})$.} Here we empirically define the window with the default value $w=2$, however, it is possible to modify this parameter during setup. After that, we compute the discrete derivative of $F'$ and sum the normalized values as follows:

\vspace{-3pt}
\begin{equation}
    \label{eq:diffMA}
    tr = \sum_{{e'}_y \in {F'}} \frac{{e'}_y - {e'}_{y-1}}{|{e'}_y - {e'}_{y-1}|}.
\end{equation}

Based on that, the trend type is defined as:

\begin{equation}
    \label{eq:pattern1}
    trend = \left\{
    \begin{array}{ll}
        \text{ascending},  & tr > 0           \\
        \text{descending}, & tr < 0           \\
        \text{neutral},    & \text{otherwise} \\
    \end{array}
    \right.\text{.}
\end{equation}

To define the pattern type $tp$ of the selected finding, we use peak detection methods which attempt to identify a local maximum by comparing neighboring values~\cite{virtanen2020scipy, Yang2009ComparisonAnalysis}. In this process, we identify all $k_p$ potential peaks within the selection, and similarly, we identify all $k_v$ potential valleys by inverting the selection data points as $F_v=-F=-I^d_c(y_a, y_b)$. Using this method, we assign a pattern factor $pf$ to the selection $F$, computing the following equation to determine whether it contains a peak, a valley, or neither.

\begin{equation}
    \label{eq:peaksfw}
    \begin{array}{ll}
        pf = |F| \times \frac{(w^+ - w^-)}{\sigma(F)}\text{, } \\
        w^+ = \sum_{k_p}W(k_p).Pr(k_p), k_p \in peaks\text{, } \\ 
        w^- = \sum_{k_v}W(k_n).Pr(k_v), k_v \in valleys\text{. } \\ 
    \end{array}
\end{equation}

In  Equation~(\ref{eq:peaksfw}), $W$ represents the width and $Pr$ the prominence, or absolute height, of the peak or valley. Therefore $w^+$ is equivalent to a probability of the selection to be a peak and $w^-$ a similar probability of being a valley. Note that $\sigma(F)$ is the standard deviation of $F=I^d_c(y_a, y_b)$. Based on the pattern factor $pf$, the type of pattern is defined as:

\begin{equation}
    \label{eq:seconddegree}
    pattern = \left\{
    \begin{array}{ll}
        \text{stable}, & \sigma(F^r_c) < \lambda_1\\
        \text{peak},     & \sigma(F) > \lambda_1 \text{ and } pf > \lambda_2                    \\
        \text{valley},   & \sigma(F) > \lambda_1 \text{ and } pf < -\lambda_2                   \\
        \text{unstable}, & \sigma(F) > \lambda_1 \text{ and } \lambda_2 \geq pf \geq -\lambda_2
    \end{array}
    \right.\text{, }
\end{equation}
where $\lambda_1$ is a threshold of whether to consider if a selection contains a pattern or not, and, if a pattern is detected, $\lambda_2$ is a threshold to define whether the selection is a peak, a valley, or an unstable oscillation. Empirically we set $\lambda_1 = 0.5$ and $\lambda_2 = 1.5$, but these parameters can be changed.

The result of this process is a pair of identifiers ($trend$ and $pattern$), which describes the selection and is exemplified in Figure~\ref{fig:patternpossibilitiesoverview}b. Finally, we define the conversion's output by using the identifier pair as keywords to be converted with the same GloVe model presented in Section~\ref{sec:indicatortags}, in which Q4EDA expands a keyword into a valid query output to be used within the output SQ.

\subsection{Query Combiner and Output Formatter}

So far, we exemplified the conversion process given a singular input to each. Each conversion outputs a single query output in the sub-expression format of Section~\ref{sec:queryspec}, with a valid sub-expression representing that specific. However, Q4EDA is not limited to only one input per data-type. Instead, each conversion process is performed for \textbf{each occurrence of its input}, 
 therefore if the input has multiple datasets $\{D1, D2, \eta\}$, the keyword conversion will be executed for each individual dataset name and return one sub-expression per dataset $T^{D1}, T^{D2}, \eta$. All other processors would similarly be executed multiple times if the VQ includes multiple occurrences of the input metadata types. 

Q4EDA then combines every sub-expression into a full expression by first creating a full combinatory permutation within each sub-expression output excluding the time-series pattern conversion, since its output will already be associated with the combination of the others. For instance, if the VQ contains two countries, one dataset name, and two ranges of years, then Q4EDA would also expect four finding sub-expressions, one for each of the country/name/years combination. With this, each of the two countries would be converted individually $T^C_{c1}$ and $T^C_{c2}$, the dataset name would be converted to $T^D_{d}$, the two-year ranges of the selection would output $T^E_{y_a, y_b}$ and $T^E_{y_c, y_d}$, and the four findings would output four distinct $T^P$. Q4EDA then expands the combinatory permutations as four inner queries by the combinatory permutations one by one and concatenates each with an ``and'' $\&$ operation, resulting in the following four sets of sub-expressions: $(T^C_{c1} \& T^D_d \& T^E_{y_a, y_b} \& T^P_{c1, d, y_a, y_b})$,  $(T^C_{c1} \& T^D_d \& T^E_{y_c, y_d} \& T^P_{c1, d, y_c, y_d})$,  $(T^C_{c2} \& T^D_d \& T^E_{y_a, y_b} \& T^P_{c2, d, y_a, y_b})$ and $(T^C_{c2} \& T^D_d \& T^E_{y_c, y_d} \& T^P_{c2, d, y_d, y_c})$. 

Then, to unite the sub-expression sets, two expressions are calculated: a full expression intersection $T_I$ by concatenating all sub-expressions sets with ``and'' $\&$ operands and a full expression union $T_U$ by concatenating all sub-expression sets with ``or'' $\|$ operands. These two are then united as $T = (T_I)^2 \| T_U$ where the full intersection is given higher weight over the full union. Note that $(T_I)^2$ applies a weight operation to a sub-expression set instead of an inner-term, as defined by our grammar. Therefore, every term within the full expression intersection should have its weight multiplied by $2$. Finally, the two full expressions are summarized through a Boolean Algebra~\cite{lashkari2009boolean} which includes Exponentiation Algebra to also solve for the weight-factor calculation.

Finally, we use an output formatter to reformat the final query calculated above into a valid SQ of a given SE. Q4EDA provides one such formatter which converts the query to the Elasticsearch simple query format~\cite{SimpleElastic} by replacing the following aspects of the query:

\begin{table}[H]
\setlength{\cellWidtha}{\textwidth/3-2\tabcolsep-0in}
\setlength{\cellWidthb}{\textwidth/3-2\tabcolsep-0in}
\setlength{\cellWidthc}{\textwidth/3-2\tabcolsep+0in}
\scalebox{1}[1]{\begin{tabularx}{\textwidth}{>{\raggedright\arraybackslash}m{\cellWidtha}>{\raggedright\arraybackslash}m{\cellWidthb}>{\raggedright\arraybackslash}m{\cellWidthc}}
\toprule
\textbf{bnf} & \textbf{Inner Example} & \textbf{Elasticsearch Equivalent} \\ \midrule
weight-factor & superscript & $\wedge$  \\
negative-factor & $-$ & N/A \\
and & $\&$ & $+$ \\
required & ``spaced term'' & +(spaced term) \\
spaced-term & some words & ``some words'' \\
\bottomrule
\end{tabularx}}
\end{table}
\noindent Therefore, other than the direct equivalents listed above, which are simply replaced, the \textit{negative-factor} is not available and removed. Note that the term itself with a \textit{negative-factor} is kept as regular \textit{operands} (e.g., ``$(-mexico^{0.5})$'' becomes ``$mexico^{0.5}$'') since querying for antonym terms of the finding can also represent information about it~\cite{zhang2009concept}.

\subsection{Query Suggestion}\label{sec:correlation}

In parallel to the SQ conversion, Q4EDA also provides query suggestions by utilizing and analyzing the available dataset collection, the VQ and the final text documents results from the conversion process, as is shown in Figure~\ref{fig:overview}. In other words, Q4EDA not only implements query expansion (\textit{R2.a}) by aggregating relevant terms to the SQ through the conversion processes but also implements query suggestions (\textit{R2.b}) in terms of the visual selection query by suggesting other partitions of the dataset collection which may potentially be related. By loosely following information retrieval techniques~\cite{ooi2015survey}, Q4EDA tackles two suggestion approaches where users are suggested related dataset names and categorical descriptors. First it provides suggestions given the presence of said dataset names or categorical descriptors within the text documents which were retrieved from the SE after executing the final SQ. Second, Q4EDA provides suggestions based on the numerical or pattern similarity of the VQ finding to other possible VQs among all the datasets and categorical descriptors.

\subsubsection{Nominal Suggestions from Text}
\label{sec:NST}

The first suggestion approach attempts to search the available nominal data's presence, which includes both the dataset name and the categorical descriptor, in a given text document. For that, a VQ is required to be converted and its output SQ executed within the target SE. The resulting text document(s) are used by this suggestion process to rank the nominal data present on each text document by number of occurrences. The result provides the suggestion of other datasets or categorical descriptors which are related to the text document being read. In order to broaden the use of this suggestion approach, we provide three implementations: direct counting, indirect counting and natural language processing (NLP), and the result is a score list which associates each nominal data to each of the texts. That is, if we follow the previous sections' examples, Q4EDA suggests the most relevant datasets and countries within UNData that relates to any given Wikipedia text document which was retrieved from Elasticsearch after executing the result of a VQ~conversion.

In the direct mode, we use a simple case-insensitive number of occurrences of all possible nominal data from the dataset, or, in other words, we apply a bag-of-words technique. That is, Q4EDA suggests a ranked list indicating the number of occurrences of every dataset name and another ranked list for the categorical descriptor. In the indirect mode, however, we first use the same keyword conversion approach of Section~\ref{sec:indicatortags} to expand the available keywords from the dataset name and categorical descriptor, and their query output's terms are then searched within the text document using bag-of-words. 
Finally, with the NLP mode, we compare the keyword conversion outputs to the extracted keyword list from the text document through another NLP method called which Gensim Keywords~\cite{rehurek2011gensim} which transforms text documents into keywords. Finally, we once again use NLTK with the GloVe model~\cite{pennington2014glove} to transform both sets of terms, namely the keyword conversion and from the text keyword extraction, into two vector embeddings which are then compared with cosine {similarity}. The result is, once again, one ranked list per nominal data which indicates suggestions indicating the similarity between the text contents and the available data within the dataset collection under analysis.

To exemplify our approach, if we assume Q4EDA setup with Elasticsearch's Wikipedia Dump~\cite{cirrussearch}, UNData~\cite{undata} and Gapminder line-charts~\cite{gapminderusaleviz}, and at some point, the user performs a VQ which, after converted and executed, returns a Wikipedia document about USA's life expectancy, then we scan the whole UNData dataset collection across the two nominal data, namely dataset name and country, and search for all their keywords (e.g., United States or Life Expectancy) within the text document. In the direct mode, we would provide two lists containing the number of occurrences of each country and indicator within the text documents. In indirect mode, we convert each dataset name and country using the keyword conversion process and search the resulting terms (e.g., United States, USA, American for the united states country or Life, Expectancy, Death for the life expectancy indicator) within the text document, and the occurrences of each term is aggregated as its respective dataset name or country as an average. The results are similarly two lists indicating the average occurrences of related terms of each country and dataset name. Finally, with the NLP mode, we also convert the text document itself into a keyword list and by converting both this list and the keyword conversion query output terms to a vector embedding format, we calculate their cosine similarity. Once again, the results are two lists indicating a similarity score between the text's generated keywords and the related terms of each country and dataset name. Independent of the mode used, the resulting lists are a country ranking list of the most related countries to the text and an indicator list of the most related indicators to the text documents.

While the direct mode provides suggestions of other nominal data present in the text and the indirect mode provides suggestions of other nominal data whose concept is present in the text, the NLP mode provides suggestions of nominal metadata whose concept matches the most prominent concepts within the text. In other words, the direct mode is better to find texts literally talking about ``life expectancy'', the indirect mode is better to find texts with information that talks about the concepts surrounding ``life expectancy'' of a country (e.g., including ``death'' and ``mortality''), and the NLP mode is better to find texts whose overall context is regarding the concepts surrounding ``life expectancy'' of a country. In the case of suggesting UNData nominal data from Wikipedia text documents, through our preliminary tests we gathered that the direct mode provided better ways to suggest nominal data based on the presence of specific words within the text, such as ``child'' of the child mortality indicator, the indirect mode provided better ways to suggest similar countries of a given VQ, and the NLP mode provided better ways to suggest similar datasets of a given VQ.

\subsubsection{Nominal Suggestions from Pattern} 
\label{sec:CSP}

\begin{figure}[H]
    \includegraphics[width=0.9\textwidth]{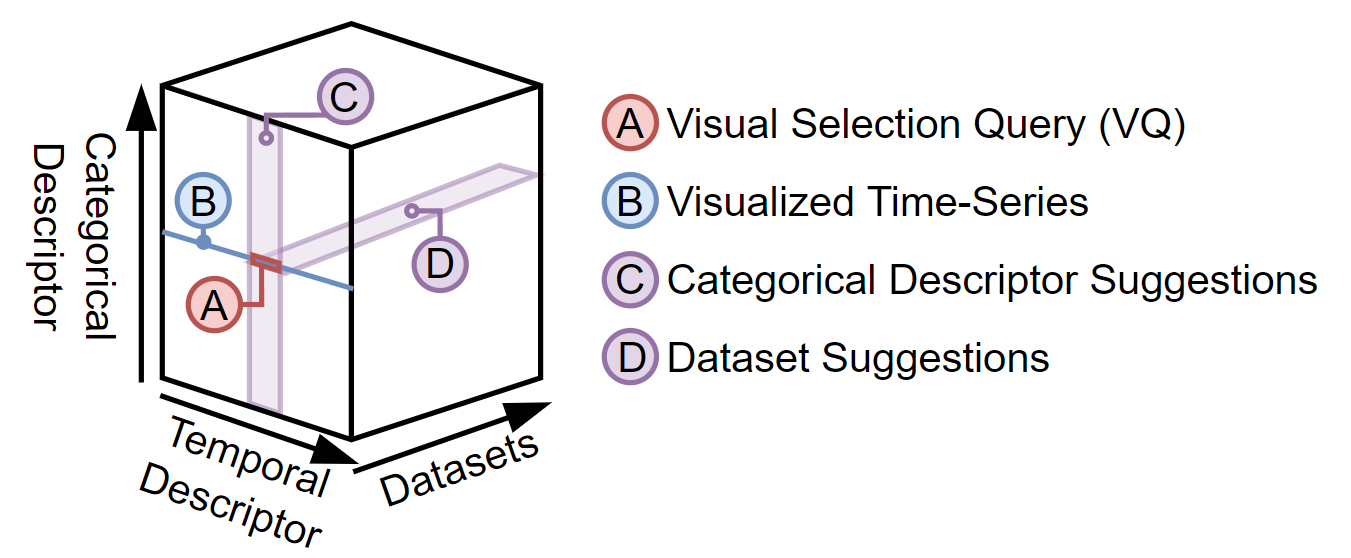}
    \caption{Example of the two suggestion lists given the dataset collection and the user's VQ (\textbf{A}) of the visualized time-series (\textbf{B}). Q4EDA calculates and outputs two suggestion lists: a dataset list indicating similar datasets to the VQ given the same time-range and categorical descriptor, and another list indicating similar categorical descriptors (e.g., country) to the VQ given the same time-range and~dataset.}
    \label{fig:findingsuggestions}
\end{figure}

The other query suggestion approach focuses on analyzing the selected finding of Section~\ref{sec:patterndetection} and providing related nominal data, such as dataset name or categorical descriptor, as suggestions for further analysis due to their similarity or dissimilarity. By comparing the time-series numeric values of the finding $F=I^d_c(y_a, y_b)=\{e_{y_a}, \ldots, e_{y_b}\}$ to all other equivalent findings among the nominal data present within the dataset collection, we are able to calculate a similarity list for each nominal data using statistical correlation techniques, as is exemplified in Figure~\ref{fig:findingsuggestions}. For instance, if considering the UNData, we would output one list indicating the similarity of the finding of all other countries $c' \in C, c' \notin c$ if all other inputs are the same, which includes the year range and dataset name, and another similarity list of all other datasets $d' \in D, d' \notin d$ if all other inputs remain the same, this time including the year range and country. Among the two lists, the most similar are suggested as related to the VQ, while the most dissimilar (e.g., lowest similarity score) are suggested as examples of an ``opposite'' or ``inverted'' pattern to the VQ.

\vspace{-6pt}

For this, we provide two different strategies to compute the aforementioned similarity: Pearson correlation and Dynamic Time-Warping (DTW)~\cite{keogh2005exact, muller2007dynamic}. These two techniques are provided because each one solves the limitations of the other. For instance, Pearson Correlation is the choice if one prefers to compare findings purely according to their general visual pattern similarity while not focusing on the distance between the selection's individual values. However, if one wishes to compare findings according to differences in amplitude or consider their raw value distances, DTW is the option. The Pearson correlation is calculated as follows:

\begin{equation}
    \label{eq:crosscorrelation}
    corr(F, F') = \frac{1}{N^2}\sum_i^N\frac{(x_i-\mu(F))({x'}_i-\mu(F'))}{\sigma(F) \sigma(F')}\text{, }
\end{equation}
where $\mu(\cdot)$ is the average value, $\sigma(\cdot)$ is the standard deviation, and $N$ is the number of values in the patterns. Correlation $corr(F, F')$ ranges in $[-1,1]$. Positive values indicate linear related series, negative inversely related series, or no relationship otherwise. 

The second option, the DTW, is a robust dissimilarity measure that finds the non-linear alignment that has the lowest accumulative Euclidean distances between points, resulting in an optimal shape match preserving magnitude~\cite{keogh2005exact, muller2007dynamic}. Since the correlation is a similarity and the DTW is an unbounded dissimilarity, we transform the DTW dissimilarity into similarity to keep consistency as follows:

\begin{equation}
    \label{eq:inverse}
    dtw_{sim}(F, F') = \dfrac{1}{1 + dtw(F, F')}\text{, }
\end{equation}
where $dtw(F, F')$ is the DTW distance between two patterns, and the resulting similarity $dtw_{sim}(F, F')$ ranges in $[0,1]$. 

The resulting suggestion lists contain the correlation score of all other possible variations for each of the respective nominal data, as seen in Figure~\ref{fig:findingsuggestions}. This list represents suggestions of similar or dissimilar findings extracted from the dataset collection given the user's VQ. With these suggestion lists, users can, for instance, directly analyze other related countries or datasets of the UNData or even visualize the lists as a similarity heat-map, or, in the case of geographic information, choropleth geography maps, for example.

\section{Use-Cases and Results}
\label{sec:results}

\subsection{UNData Line Charts and Wikipedia}
\label{sec:usagelinechart}

This section presents a hypothetical usage scenario of Q4EDA, showing how it can assist external VA tool users in understanding patterns of interest and building up extra knowledge during exploratory data analysis. Here we use a replica of Gapminder's line chart visualization~\cite{gapminderusaleviz} to explore the UNData dataset~\cite{undata} as a VA tool couple with an Elasticsearch~\cite{elasticsearchwikipedia} SE loaded with Wikipedia text documents~\cite{cirrussearch}. 

\begin{figure}[H]

\begin{adjustwidth}{-\extralength}{0cm}
\centering 
    \includegraphics[width=\linewidth]{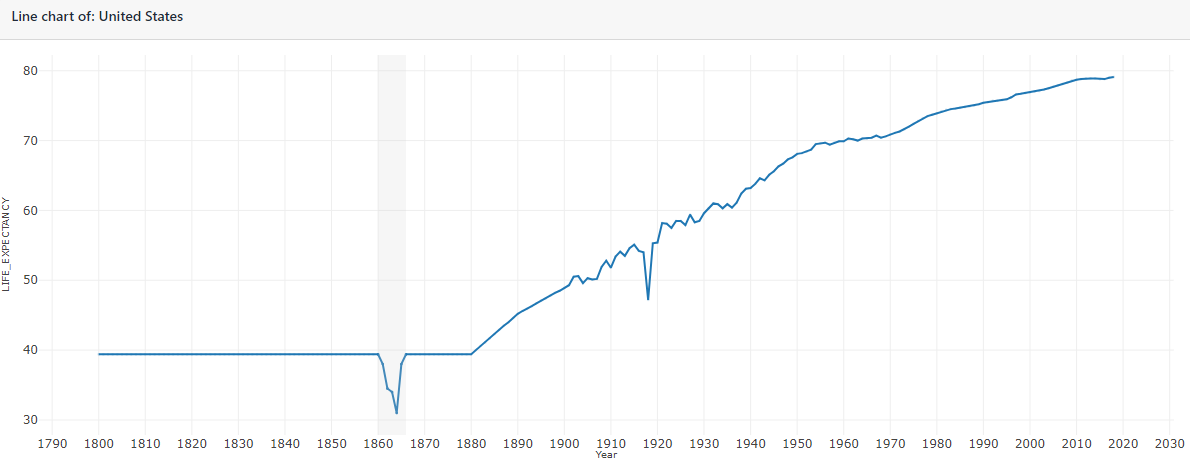}
\end{adjustwidth}
    \caption{Life expectancy line chart of the United States. Two noticeable valleys are observed between $1860$ and $1866$ and between $1917$ and $1919$. The gray area represents the user selection.}
    \label{fig:usalifeexpectancyline}
\end{figure}

We introduce Justin, our fictional American High School student. He wants to investigate how the average lifespan has changed over the years in the United States. With that in mind, he visualizes the ``Life Expectancy'' indicator of the ``United States'' country (Figure~\ref{fig:usalifeexpectancyline}). The resulting line-chart shows a positive trend toward increasing the overall American lifespan over the years. However, he notices two interesting patterns, one valley between $1860$ and $1866$ and another between $1917$ and $1919$ with some instability between $1901$ and $1930$.

\begin{figure}[H]

\begin{adjustwidth}{-\extralength}{0cm}
\centering 
    \subfloat[\centering] {\includegraphics[width=7.2cm]{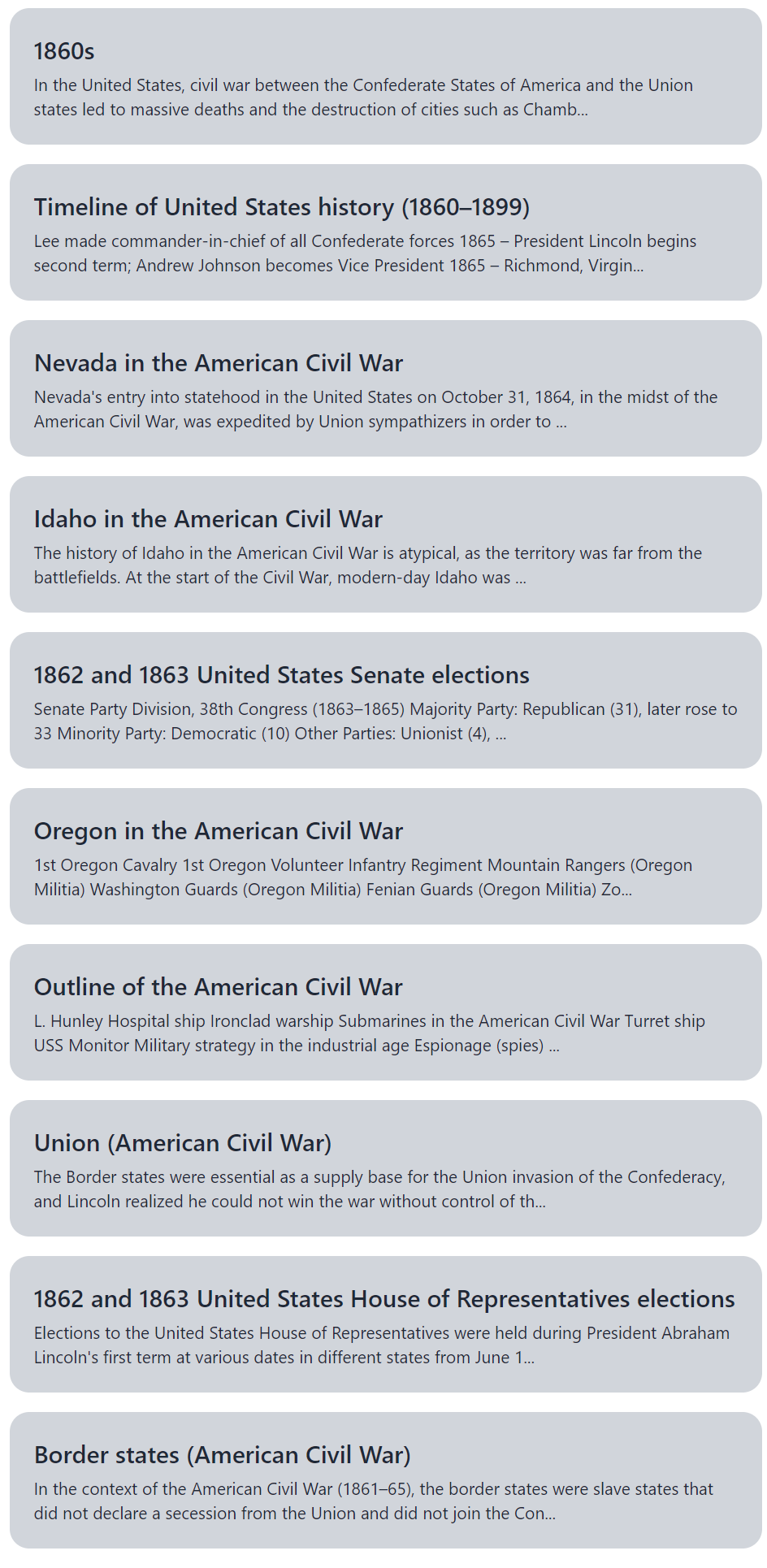}\label{fig:usalifeexpectancydocs1}}\quad
    \subfloat[\centering] {\includegraphics[width=7.2cm]{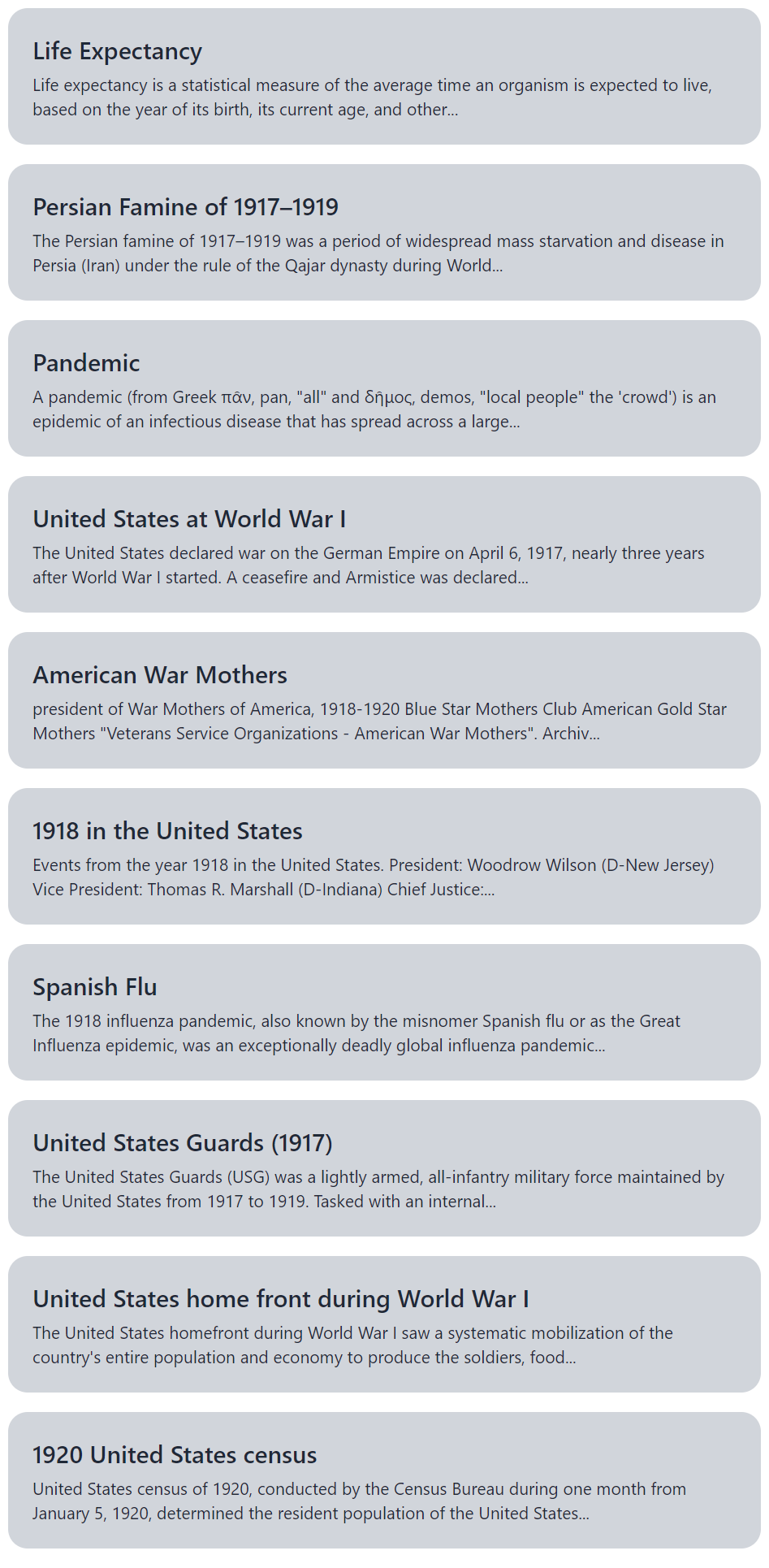}\label{fig:usalifeexpectancydocs2}}
\end{adjustwidth}
    \caption{List of  
documents retrieved from Wikipedia related to the drop in the United States' life expectancy. (\textbf{a}) Selection between $1860$ and $1866$. (\textbf{b}) Selection between $1917$ and $1919$.}
    \label{fig:usalifeexpectancydocs}
\end{figure}

To further inspect these patterns and understand what is happening, Justin first selects the left-most valley (between $1860$ and $1866$). Internally, the enhanced VA tool uses Q4EDA to generate a search query, retrieves related Wikipedia documents, and gets a list of suggestions of related countries and datasets considering the selection (Section~\ref{sec:correlation}). Based on that, Justin adds the two top-ranked suggested countries to the line chart (Sweden and the United Kingdom) and realizes that the drop in life expectancy is probably an American effect since even the most related countries do not present similar valley in the same period (omitted due to space constraints). With that in mind, Justin checks the retrieved documents and observes the prevalence of the terms ``civil war'' in the returned snippets (Figure~\ref{fig:usalifeexpectancydocs}a) and concludes that the lower life expectancy rating may result from a civil war. By reading some of the retrieved documents, he learns that the trigger of the civil war was the result of Abraham Lincoln's election and the United States southern states feeling unrepresented and/or challenged due to slavery, being able to discover one important piece for the storytelling of the United States lifespan variations, including its~trigger.

\vspace{-12pt}

\begin{figure}[h]

\begin{adjustwidth}{-\extralength}{0cm}
\centering 
    \includegraphics[width=\linewidth]{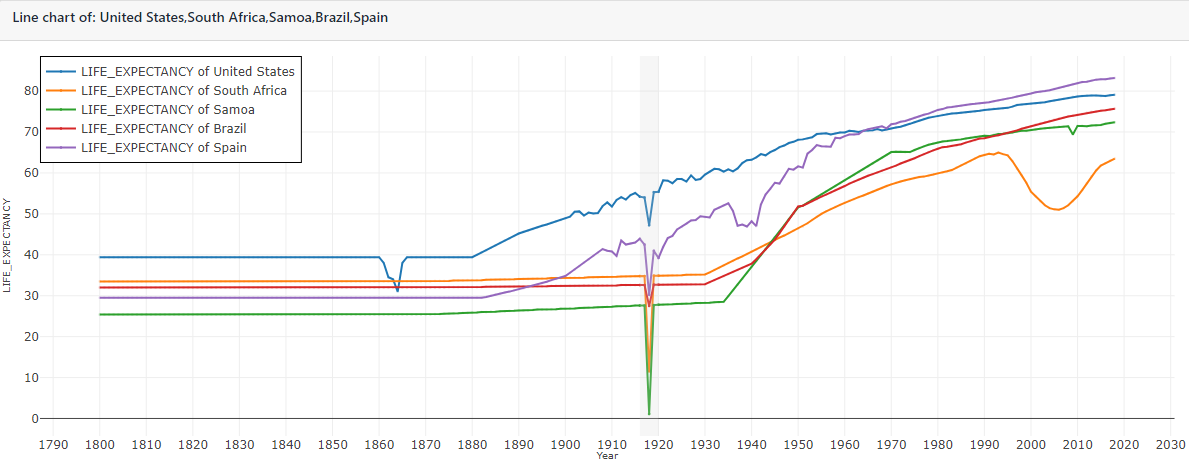}
\end{adjustwidth}
    \caption{Life expectancy line chart multiple countries. All countries present the same valley between $1917$ and $1919$, indicating a global reason for the drop in life expectancy.}
    \label{fig:usalifeexpectancylineothersef}
\end{figure}

Justin follows up to investigate the second dip in life expectancy and selects the period between $1917$ and $1919$. The retrieved documents' contents are, however, less homogeneous with different causes for the drop, with a span of documents discussing different topics (Figure~\ref{fig:usalifeexpectancydocs}b). As in the previous example, he adds other top-ranked suggested countries to the line chart, including countries from different continents. Differently from the previous valley, all the included countries present a similar drop in life expectancy in the same period (Figure~\ref{fig:usalifeexpectancylineothersef}), apparently suggesting that a global event took place. Rechecking the retrieved documents, he infers that some potential reasons for the changes in life expectancy may be World War I and the Spanish Flu, also called Influenza Pandemic.

\begin{figure}[H]

\begin{adjustwidth}{-\extralength}{0cm}
\centering 
    \subfloat[\centering] {\includegraphics[width=8.6cm]{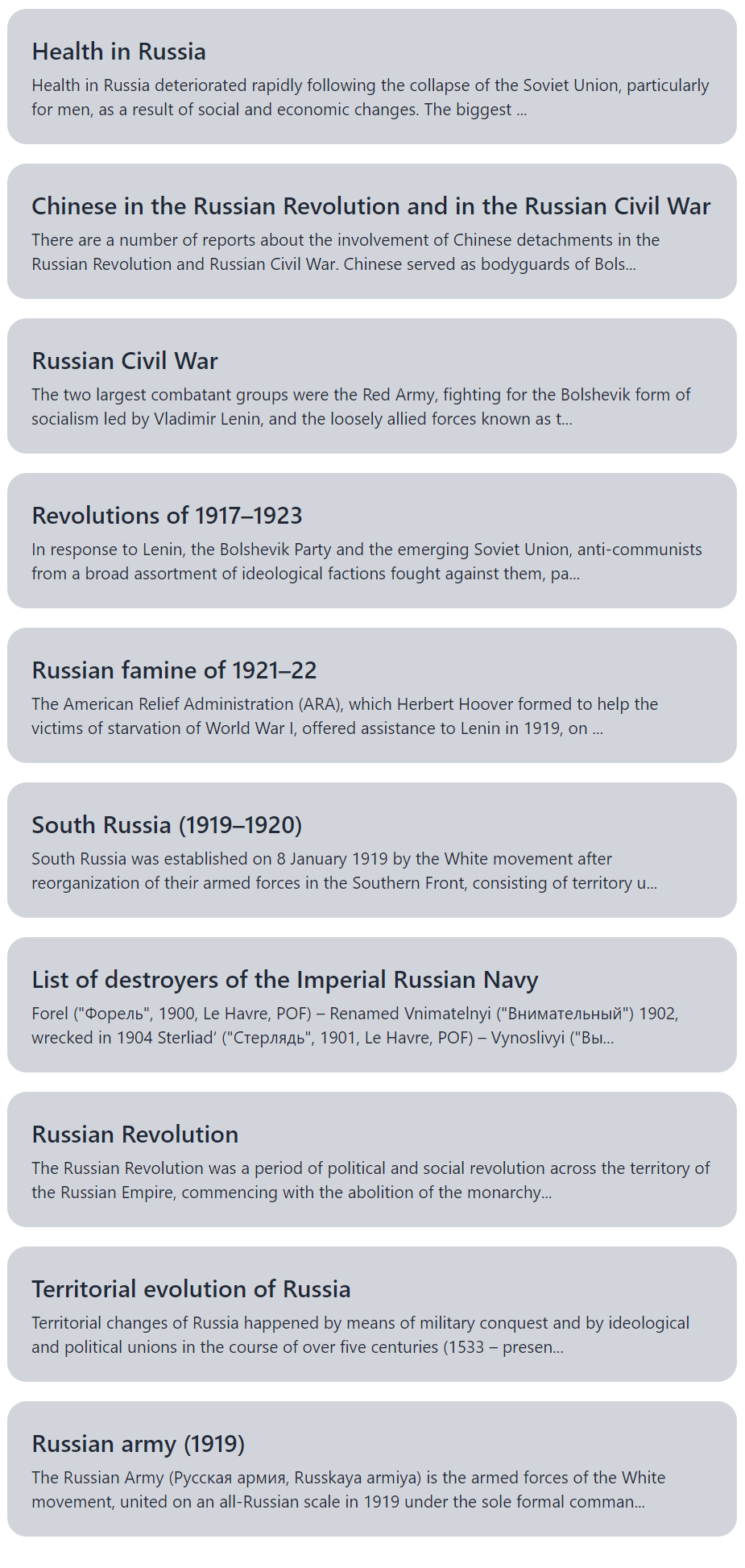}\label{fig:ursslifeexpectancydocs}}\quad
    \subfloat[\centering] {\includegraphics[width=8.6cm]{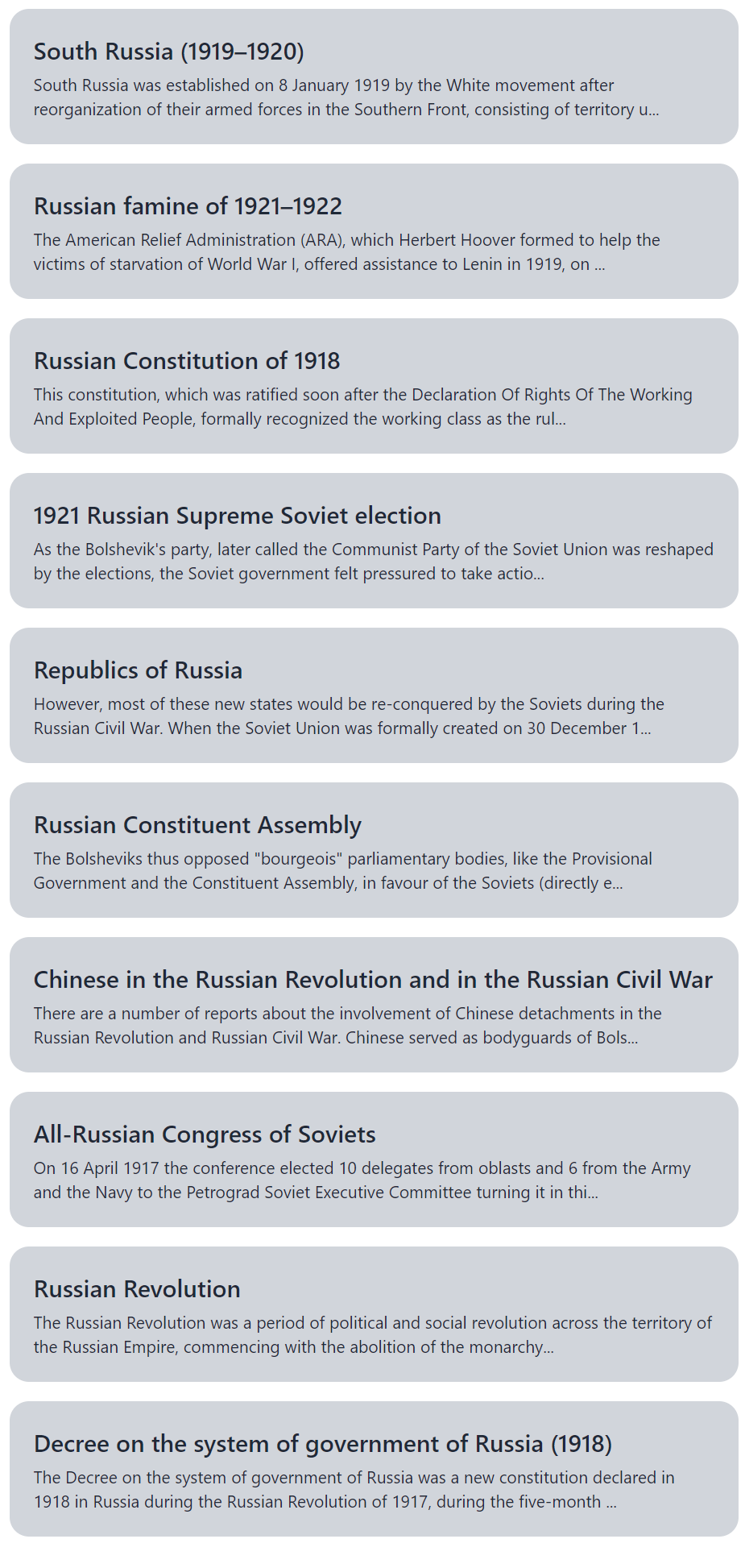}\label{fig:ursslifeexpectancydemocracydocs}}
\end{adjustwidth}
    \caption{List 
 of documents retrieved from Wikipedia related to the pattern in the Russia's dataset between $1917$ and $1919$, indicating several potential events that may have affected it. (\textbf{a}) Selection of the valley of Russia's life expectancy. (\textbf{b}) Selection of the drop of Russia's democracy index score.}
    \label{fig:ursswiki}
\end{figure}

Still using the countries suggestion list, Justin discovers that Russia is amongst the lowest-ranked countries and decides to investigate---here we are discussing the country suggestion as a ranked list. Adding Russia to the line chart, Justin finds out that the reason for the similarity score being low in this period is that Russia's valley is much wider than the United States (image omitted due to space constraints). 
By selecting the valley in Russia's life expectancy and checking the suggested countries, he discovers that Belarus, Ukraine, Turkmenistan, Uzbekistan, Tajikistan, Kazakhstan, and Uzbekistan are the most similar to Russia. By checking the retrieved documents (Figure~\ref{fig:ursswiki}a), Justin discovers that during $1917$ and $1919$, Russia was facing the abolition of its monarchy in $1917$, a civil war, and the beginning of the Soviet Union, besides likely facing World War I and the Influenza Pandemic as he saw with the rest of the world.

\begin{figure}[H]

\begin{adjustwidth}{-\extralength}{0cm}
\centering 
    \includegraphics[width=\linewidth]{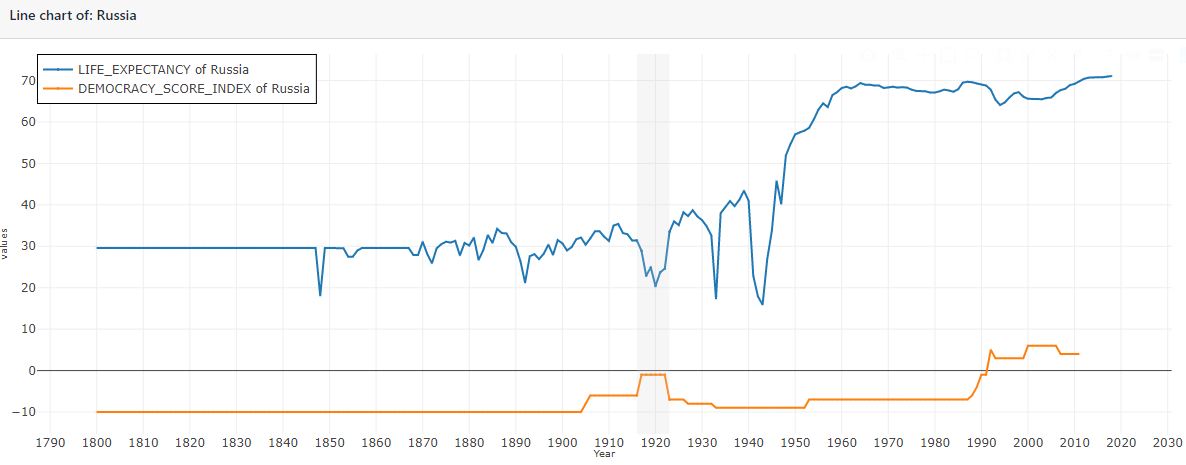}
\end{adjustwidth}
    \caption{Russia life expectancy and democracy index presents interesting similarities.}
    \label{fig:ursslifeexpectancydemocracy}
\end{figure}

Besides suggesting countries, Q4EDA also suggests other datasets with correlations given a selection (Section~\ref{sec:CSP}). Justin notices that the “Democracy Index” dataset is suggested, so he creates a line-chart displaying Russia's life expectancy and democracy index together (Figure~\ref{fig:ursslifeexpectancydemocracy}). In the resulting visual representation, he notices an inverted pattern (a valley in the life expectancy and a peak in the democracy index) between $1917$ and $1923$. Justin then selects such time period and, considering the list of returned Wikipedia documents (Figure~\ref{fig:ursslifeexpectancydemocracydocs}), he discovers that the sudden change in the democracy index of Russia can be attributed in some part to Russia's constitution of $1918$, “Russian Famine”, “Russia's Civil War” and the “Russian Revolution”, which tells about how the monarchy was abolished in $1917$ and the Soviet Union was established in $1923$, matching the steep fall of Russia's democracy index in $1923$. Justin concludes that the political landscape variation in Russia definitively impacted the lives of the Russian population, including their life~expectancy.

Although an elementary analysis, Justin discovered pieces for the life-expectancy variations in the United States, including its trigger. Two different global situations, World War I and the Spanish Flu. Why Russia was differently affected than the rest of the world at the time, and much of Russia's history on revolutions and economic crisis.

\vspace{-12pt}

\subsection{Inner Query Stability Evaluation}
\label{sec:querystability}

Although so far we assumed precise visual queries within our examples, user selection is expected to not be exact every time due to the inherent limitations of visual interfaces and means of visual selection. Therefore, we measure how much the Wikipedia results change when slight variations of VQ occur in order to evaluate \textit{R1.b} (see Section~\ref{sec:design}). For this, we follow \cite{Memon2004WhatTesting} and create an oracle that emulates the user behavior, allowing us to test this aspect of Q4EDA exhaustively. Notice that in this evaluation, we are not validating the search results themselves since we could not find a labeled dataset that would allow us to evaluate the connection between time-series visual patterns and the text documents of a search engine such as Elasticsearch.

The oracle implementation we use is straightforward. Given a time series, we iterate over it, automatically extracting the top patterns classified by the height of their valleys and peaks; then, we vary a window around each of the patterns to generate similar selections that emulate users' selection inaccuracies. In our experiments, the window size is set to $3$, resulting in $9$ queries per pattern (one original and eight derived). This value was defined by running a test with our lab members asking them to manually select patterns in several different time series, considering the average variation among them to set the window. The stability of our query process is then measured by comparing the intersection of the set of documents retrieved using the original query $q$ and the sets of documents fetched using the derived queries $q'$. Given $D$ the set documents retrieved using the original query $q$ and $D_i$ the list of documents returned using one of the derived queries $q'_i$, the stability is computed~as:

\begin{equation}
\frac{1}{|D| \cdot |\Delta|} \sum_{D_i \in \Delta} |(D \cap D_i)|\text{, }
\end{equation}
where $\Delta$ is the set containing the lists of documents produced by all derived queries $q'$. Notice that the number of documents in $D$ and $D_i$ is the same and defined by the number of documents we display in our interface. In our tests, we set it to $10$.

\begin{figure}[H]
    \includegraphics[width=1\columnwidth]{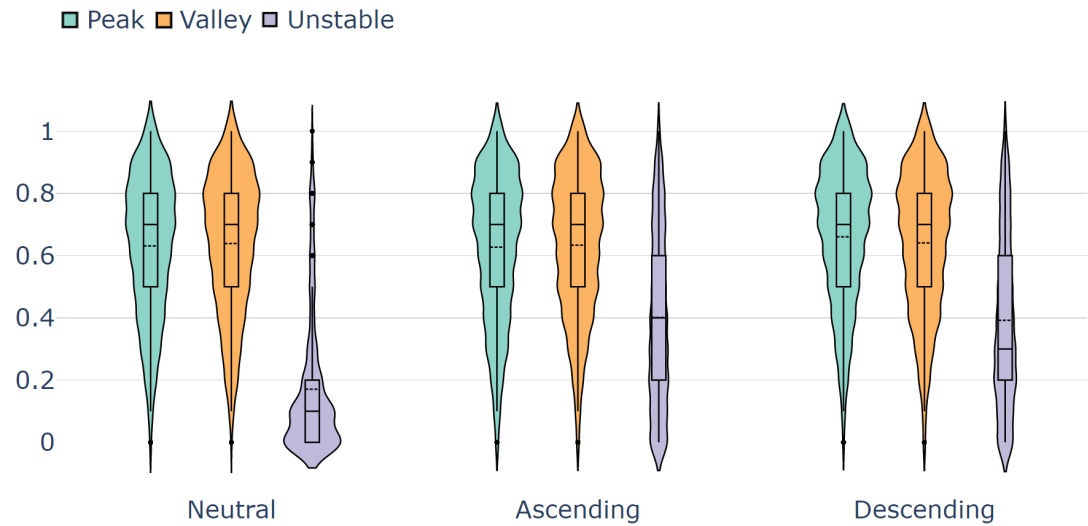}
    \caption{Query stability analysis. On average, slight variations in the selection return $39\%$ to $64\%$ of the documents returned by the original query depending on the pattern type, indicating a good degree of stability and reproducibility, especially for patterns with peaks and valleys.}
    \label{fig:querystability}
\end{figure}

To execute a comprehensive test, we select $960$ time series from the UNData dataset~\cite{undata} and measure the query stability for each one. We automatically detected 5286 relevant patterns, resulting in $47,574$ queries submitted to the search engine. Figure~\ref{fig:querystability} summarizes the results. Overall, on average, the stability is $0.5121$, meaning that slight variations in the selection return $51\%$ of the documents returned by the original query. More specifically, peaks and valleys are more stable, with an average of around $64\%$ and a standard deviation of $0.22$. At the same time, 'unstable patterns' have less document stability with an average of $39\%$ and a standard deviation of $0.27$. This suggests that the peaks and valleys are meaningful within Q4EDA, indicating that it appropriately translates well-defined visual patterns into coherent groups of documents. Although this cannot quantify the quality of the retrieved documents, it indicates that users with similar behavior are expected to receive similar documents when attempting to select the same pattern, indicating a good degree of stability and reproducibility.

\vspace{-6pt}
\section{User Evaluation}
\label{sec:usereval}

Using the same workflow as Justin's usage scenario (see Section~\ref{sec:usagelinechart}), we conducted a user study to evaluate: (1) whether Q4EDA conversion method allows users to more accurately find textual information related to a specific time series pattern if compared to manually searching (\textit{R1} and \textit{R2.a}); and (2) whether users are more accurate in their query results using Q4EDA even when confident in their findings (\textit{R1.a}). We use a replica of Gapminder~\cite{rosling2012gapminderorg, gapminderusaleviz} line-chart to represent the times-series and the Wikipedia search engine. We use this setup given the Gapminder popularity and to avoid search results variations that may occur between different user profiles if, for instance, the Google search engine is employed. Through this study, we aim to check the following null hypotheses: 

\begin{quote}
$\mathbf{H^P_0}$:  
There is no difference in the amount of correct information related to a given pattern of interest a user can find using the proposed visual query conversion and manually querying the target search engine.
\end{quote}

\begin{quote}
$\mathbf{H^C_0}$: There is no difference in the amount of correct information related to a given pattern of interest a user who was \textbf{confident} in their answer can find using the proposed visual query conversion and manually querying the target search~engine.

\end{quote}

We recruited $21$ participants aged $16$ and up. All but two are from computer sciences, one is from social sciences, and the other claim to not have a primary area of study. Due to COVID-19, we were not able to conduct an in-person study. Instead, we conducted it through a self-guided online survey. Our experiment first conducts a demographic questionnaire and introduces the Gapminder replica with a video tutorial. Participants are then guided through an interactive tutorial using the system and finally are given a list of tasks to be performed. The study was executed in an automated and non-obstructive manner through an online survey system. Participants joined the study through their machines and had unrestricted time to complete it. The study took on average $50$ min, but due to the self-guided nature of online surveys, a third of the participants did not complete the study in one continuous session. 

To evaluate our hypotheses, we randomly split our participants into two groups, one containing $11$ participants, which we call \textit{Cohort 1} (\textit{C1}), and another containing $10$~participants, which we call \textit{Cohort 2} (\textit{C2}). Both groups execute two sets of tasks. For the first task, \textit{C2} is the experimental group while executing tasks using Q4EDA, and \textit{C1} is the control group executing tasks without visual queries but instead manually searching Wikipedia to find related information to the observed patterns. Therefore no search query conversion capabilities are provided to \textit{C1} as opposed to \textit{C2}. For the second task, the roles of \textit{C1} and \textit{C2} are swapped.

The benchmark of this user study was performed by manually searching both Wikipedia and through visual query for the expected text documents and phrases and confirming that all tasks can be performed with both tools. It is important to note that both tools use and provide interfaces to the same time-series dataset and the same textual dataset. The study does not attempt to capture all possible historical events and facts related to a pattern but only to capture some equally available to be queried in both formats.

For the first set of tasks, using the ``United States life expectancy'' dataset, we asked participants to search for probable causes for the drop between $1860$ and $1866$. We presented $5$ alternatives and asked which one is related to the observed patterns (four correct and one incorrect). Answers from the experimental group using Q4EAD were better in finding more texts related to the time-series pattern. On average, participants manually searching Wikipedia (\textit{C1}) answered correctly $31\%$ of the alternatives presented regarding related causes against $48\%$ using Q4EDA (\textit{C2}). For this task, all participants but one from \textit{C2} said they now understand better Q4EDA's usefulness for this kind of task and, by using the Q4EDA's suggestions (see Section~\ref{sec:correlation} and \textit{R2.b}), $7$ participants from the experimental group were also able to find other related indicators that may further enhance the exploratory~analysis. 

In the second task, we swapped the two groups so that \textit{C1} is the experimental group using Q4EDA's query conversion and suggestions and \textit{C2} is the control group directly querying Wikipedia. In this task, we ask users to investigate another drop in the United States life expectancy between $1916$ and $1919$. Similar to the previous set of tasks, participants using Q4EDA had an overall higher performance. Participants manually searching Wikipedia (\textit{C2}) correctly answered  $38\%$ of the alternatives presented of related causes against $68\%$ using Q4EDA (\textit{C1}).

To verify our null hypothesis $\mathbf{H^P_0}$ across the two sets of tasks, we compared the control group directly using only Wikipedia and the experimental group using Q4EDA across \textbf{all} tasks. The calculated $p$-value is $P_v^{H_0}=0.0006 < 0.05$ and t-value is $T_v^{H^P_0}=-4.33$. Therefore we can confidently reject $\mathbf{H^P_0}$ and affirm that the experimental group was statistically better in finding textual information related to a pattern of interest when compared to the control group. In more specific terms, when comparing the number of answers from each group which matched our expected answers, the control group had, on average, identified $4.33$~fewer pieces of information matching our expectations than the experimental group. 

We also asked participants how confident they were with their answers to our tests. We first collected the answers where the participant was at least $Conf\geq\textit{Agree}$ in their confidence level. The resulting $p$-value is $P_v=0.011 < 0.05$, therefore rejecting the hypothesis $H^C_0$, indicating that among the participants who claim are confident in their answers, the control group has fewer answers matching what was expected when compared to similarly confident participants from the experimental group. When considering that these two groups were similarly confident in their answers, we conclude that the confidence of the control group does correlate to a correct answer as well as experimental group, or in other words, given a pattern of interest, confident Q4EDA users have more accurate textual findings than users who directly queried Wikipedia.


In conclusion, participants of the experimental group significantly outperformed participants of the control group, showing that Q4EDA is effective in providing means for users to better and more correctly discover information in regards to a visual pattern and also to provide an effective way to convert a visual query into useful Wikipedia queries when compared to constructing the query manually.

\section{Discussions and Limitations}
\label{sec:limitations}

This paper has explored a method to convert a visual selection query into a search query usable within search engines. We tested this technique by converting patterns within world indicator dataset collections and enhancing users' analysis through Wikipedia documents. Indeed, our system is open and easily usable in other domains, which may also benefit from interpreting time-series visual selections as search queries. {Different input datasets, such as news, stock market, financial data, IoT sensor data, social media, and natural disasters dataset collections could also be used in a similar fashion.} Although some of such dataset collections may be directly supported by the processes implemented so far, others may contain new data-types, which would also require new conversion processes to be implemented, requiring modification to Q4EDA's structure. We leave  the application of Q4EDA to other domains for future work.


Since we implemented Q4EDA's basic functionality, most of the exploratory tests we executed resulted in valuable retrieved textual information. However, it failed to bring meaningful documents for some specific demography indicators during our tests, though manually searching the same search engine gave similar poor results. Indeed, the usefulness of the query conversion depends on whether the SE has relevant information about the selected patterns and how well both the user and the SE can parse the data. For instance, if Wikipedia has no information on some subject for a given pattern, our framework may still return documents that may aggregate next to nothing for the analysis, which is not surprising. We forgo how to match other time-series datasets to potential SEs, expecting instead this to be decided by whoever sets up and uses our framework (e.g., external VA~tools).


Another limitation pertains to the suggestion approaches. Q4EDA suggests relevant nominal data given the visual query and the text documents by counting the number of the textual elements shared among the two either directly or through NLP. Updates to the suggestion processes are planned so that Q4EDA can also include the temporal descriptor (e.g., year) within the search or include semantic analysis over the text. The suggestion and keyword conversion processes could also benefit from other NLP techniques. Important to note that we tested other NLP techniques as part of our development process, namely Latent Dirichlet Allocation (LDA)~\cite{blei2003latent} and DistilBert~\cite{reimers-2019-sentence-bert}. We, however, landed on the ones presented due to their speed in processing hundreds of text documents in less than a minute.

%
{We also tested our usage scenario and user evaluation using Google as a search engine, and the preliminary results were arguably more informative than Wikipedia.} Indeed, Google API was our first choice for most of our examples. However, its connection to the Gapminder replica was not ideal due to Google's imposed limitations, such as the limited number of API calls and limited information parsing. Other challenges were also encountered, such as the processing of unstructured web pages as opposed to well-defined Wikipedia documents and the significant variations of results given the user's profile. 
Therefore, even though both output modules were available, we opted for Wikipedia for our examples and user evaluation. {We also observed this issue when considering our evaluation process. Seen that there is no labeled dataset where patterns within time-series data are linked to textual information within a search engine, Q4EDA was unable to use common metrics, such as accuracy or f1-score, to evaluate its resulting query. However, we hope Q4EDA will instigate such labeled datasets to be compiled in the future.}

%
%
{Finally, although our main usage scenario focuses on a line-chart view, Q4EDA does not impose such requirements on the visual metaphor. For instance, if the Gapminder's bubble chart~\cite{gapminderbubbleviz} is used, a VQ over it would consider the two datasets displayed as the chart's axis.} Hypothetically, such a VA tool could expect a box selection within the main bubble chart visualization, which would make the VQ include one year and a range of countries. However, the VA tool could also expect a selection within the animation timeline of the bubble chart, which would make the VQ include a range of years and all available countries or even a combination of the two selection modes. In summary, Q4EDA is agnostic in terms of which visual representation was chosen.
\section{Conclusions}
\label{sec:conclusions}

In this paper, we present Q4EDA, a framework that converts a \textit{visual selection query} into a \textit{search query} format to be used in existing \textit{search engines} and, from its results, suggests other potential aspects of the data to be analysed, all of which provide a novel strategy to utilize user input for textual information retrieval.
The usefulness of Q4EDA is brought by an application linking a Gapminder's line-chart replica with Wikipedia documents to support exploratory analysis of world indicators. The improvement in users' exploratory analysis capability is then confirmed through a user test showing that users can find more information using Q4EDA compared to the standard manual keyword-based queries, especially when confident in their findings. The stability of the conversion process given slight variations of its inputs was also evaluated to in order to verify the applicability of Q4EDA with inaccurate visual selection interfaces.
Despite its limitations, Q4EDA is unique in its proposal, representing an advance towards providing solutions for querying textual information from general purpose search engines based on user interaction with visual representations.

\vspace{6pt} 




\authorcontributions{
Conceptualization, L.C. and F.P.; methodology, L.C., M.F. and F.P.; software, L.C. and M.F.; validation, L.C. and F.P.; formal analysis, L.C. and F.P.; investigation, L.C. and F.P.; data curation, L.C. and M.F.; writing---original draft preparation, L.C. and F.P.; writing---review and editing, L.C. and F.P.; visualization, L.C. and F.P.; supervision, F.P; project administration, F.P. All authors have read and agreed to the published version of the manuscript.
}

\funding{This research received no external funding.}


\institutionalreview{
The study protocol was approved by the Institutional Review Board (or Ethics Committee) of Dalhousie University (REB \# 2020-5310, 22 February 2021).
}

\informedconsent{Informed consent was obtained from all subjects involved in the study under REB \#2020-5310.}

\dataavailability{Not applicable} 

\acknowledgments{Authors acknowledge the support of the Natural Sciences and Engineering Research Council of Canada (NSERC).}

\conflictsofinterest{The authors declare no conflict of interest.} 



\abbreviations{Abbreviations}{
The following abbreviations are used in this manuscript:\\

\noindent 
\begin{tabular}{@{}ll}
SE & Search Engine\\
SQ & Search Query\\
VA & Visual Analytics\\
EDA & Exploratory Data Analysis
\end{tabular}
}

\appendixtitles{no} 



\begin{adjustwidth}{-\extralength}{0cm}

\reftitle{References}

\end{adjustwidth}

\begin{thebibliography}{999}

\bibitem[Croft \em{et~al.}(2010)Croft, Metzler, and Strohman]{croft2010search}
Croft, W.B.; Metzler, D.; Strohman, T.
\newblock {\em Search Engines: Information Retrieval in Practice}; Addison-Wesley Reading: Boston, MA, USA, 2010; Volume 520.

\bibitem[sqs(accessed in 2022-05-06)]{sqsdef}
What Is a Search Query? (Definition)---Seo Glossary. 
\newblock Available online: \url{https://growhackscale.com/glossary/search-queries} (accessed on 6 February 2020).

\bibitem[goo(accessed in 2022-05-06)]{google}
Google Search.
\newblock Available online: \url{https://www.google.com/} (accessed on 6 February 2020).

\bibitem[wik(accessed in 2022-05-06)]{wikipedia}
Wikipedia---The Free Encyclopedia.  Available online: \url{https://en.wikipedia.org/wiki/Main_Page} (accessed on 6 May 2022). 

\bibitem[und(accessed in 2022-05-06)]{undata}
United Nations Datasets.
\newblock Available online: \url{https://data.un.org/} (accessed on 6 February 2020).

\bibitem[Rosling(2012)]{rosling2012gapminderorg}
Rosling, H.
\newblock Data---Gapminder.org. Available online: \url{https://www.gapminder.org/} (accessed on 6 February 2020). 


\bibitem[Sarma(2017)]{sarma2017hans}
Sarma, A.
\newblock Hans Rosling brought data to life, showed our misconceptions about
  the world.
\newblock {\em Skept. Inq.} {\bf 2017}, {\em 41},~9--10.

\bibitem[Kammerer and Bohnacker(2012)]{kammerer2012children}
Kammerer, Y.; Bohnacker, M.
\newblock Children's web search with Google: The effectiveness of natural
  language queries.
\newblock In Proceedings of the 11th International
  Conference on Interaction Design and Children, Bremen, Germany, 12--15 June 
 2012; pp. 184--187.

\bibitem[Reilly and Thompson(2017)]{reilly2017reverse}
Reilly, M.; Thompson, S.
\newblock Reverse image lookup: assessing digital library users and reuses.
\newblock {\em J. Web Librariansh.} {\bf 2017}, {\em 11},~56--68.

\bibitem[Cafarella and Etzioni(2005)]{cafarella2005search}
Cafarella, M.J.; Etzioni, O.
\newblock A search engine for natural language applications.
\newblock In Proceedings of the 14th International
  Conference on World Wide Web, Chiba, Japan, 10--14 May 2005; pp. 442--452.

\bibitem[Hullman \em{et~al.}(2013)Hullman, Diakopoulos, and
  Adar]{hullman2013contextifier}
Hullman, J.; Diakopoulos, N.; Adar, E.
\newblock Contextifier: Automatic generation of annotated stock visualizations.
\newblock In Proceedings of the SIGCHI Conference on Human
  Factors in Computing Systems, Paris, France, 27 April--2 May 2013; pp. 2707--2716.

\bibitem[Badam \em{et~al.}(2018)Badam, Liu, and Elmqvist]{badam2018elastic}
Badam, S.K.; Liu, Z.; Elmqvist, N.
\newblock Elastic documents: Coupling text and tables through contextual
  visualizations for enhanced document reading.
\newblock {\em IEEE Trans. Vis. Comput. Graph.} {\bf
  2018}, {\em 25},~661--671.

\bibitem[{Yu} and {Silva}(2020)]{yu2019flowsense}
{Yu}, B.; {Silva}, C.T.
\newblock FlowSense: A Natural Language Interface for Visual Data Exploration
  within a Dataflow System.
\newblock {\em IEEE Trans. Vis. Comput. Graph.} {\bf
  2020}, {\em 26},~1--11.

\bibitem[Kraska(2018)]{kraska2018northstar}
Kraska, T.
\newblock Northstar: An interactive data science system.
\newblock {\em Proc.  VLDB Endow.} {\bf 2018}, {\em
  11},~2150--2164.

\bibitem[Zhou \em{et~al.}(2021)Zhou, Wen, Wang, and Gotz]{zhou2021modeling}
Zhou, Z.; Wen, X.; Wang, Y.; Gotz, D.
\newblock Modeling and {Leveraging} {Analytic} {Focus} {During} {Exploratory}
  {Visual} {Analysis}.
\newblock {\em arXiv} {\bf 2021}, arXiv:2101.08856.

\bibitem[Borland \em{et~al.}(2019)Borland, Wang, Zhang, Shrestha, and
  Gotz]{borland2019selection}
Borland, D.; Wang, W.; Zhang, J.; Shrestha, J.; Gotz, D.
\newblock Selection bias tracking and detailed subset comparison for
  high-dimensional data.
\newblock {\em IEEE Trans. Vis. Comput. Graph.} {\bf
  2019}, {\em 26},~429--439.

\bibitem[Borland \em{et~al.}(2020)Borland, Zhang, Kaul, and
  Gotz]{borland2020selection}
Borland, D.; Zhang, J.; Kaul, S.; Gotz, D.
\newblock Selection-Bias-Corrected Visualization via Dynamic Reweighting.
\newblock {\em IEEE Trans. Vis. Comput. Graph.} {\bf
  2020}, \emph{27}, 1481--1491.

\bibitem[Ooi \em{et~al.}(2015)Ooi, Ma, Qin, and Liew]{ooi2015survey}
Ooi, J.; Ma, X.; Qin, H.; Liew, S.C.
\newblock A survey of query expansion, query suggestion and query refinement
  techniques.
\newblock In Proceedings of the 2015 4th International Conference on Software
  Engineering and Computer Systems (ICSECS), Kuantan, Malaysia, 19--21 August 2015; pp. 112--117.

\bibitem[Yi \em{et~al.}(2017)Yi, Choi, Bhowmick, and Xu]{yi2017autog}
Yi, P.; Choi, B.; Bhowmick, S.S.; Xu, J.
\newblock AutoG: a visual query autocompletion framework for graph databases.
\newblock {\em VLDB J.} {\bf 2017}, {\em 26},~347--372.

\bibitem[Zhang \em{et~al.}(2012)Zhang, Stoffel, Behrisch, Mittelstadt, Schreck,
  Pompl, Weber, Last, and Keim]{zhang2012visual}
Zhang, L.; Stoffel, A.; Behrisch, M.; Mittelstadt, S.; Schreck, T.; Pompl, R.;
  Weber, S.; Last, H.; Keim, D.
\newblock Visual analytics for the big data era---A comparative review of
  state-of-the-art commercial systems.
\newblock In Proceedings of the 2012 IEEE Conference on Visual Analytics
  Science and Technology (VAST), Seattle, WA, USA, 14--19 October 2012; pp. 173--182.

\bibitem[Srinivasan \em{et~al.}(2018)Srinivasan, Drucker, Endert, and
  Stasko]{srinivasan2018augmenting}
Srinivasan, A.; Drucker, S.M.; Endert, A.; Stasko, J.
\newblock Augmenting visualizations with interactive data facts to facilitate
  interpretation and communication.
\newblock {\em IEEE Trans. Vis. Comput. Graph.} {\bf
  2018}, {\em 25},~672--681.

\bibitem[Suh \em{et~al.}(2022)Suh, Jiang, Mosca, Wu, and Chang]{suh2022grammar}
Suh, A.; Jiang, Y.; Mosca, A.; Wu, E.; Chang, R.
\newblock A Grammar for Hypothesis-Driven Visual Analysis.
\newblock {\em arXiv} {\bf 2022},  arXiv:2204.14267.

\bibitem[Cui \em{et~al.}(2019)Cui, Zhang, Wang, Huang, Chen, Fang, Zhang, Lou,
  and Zhang]{cui2019text}
Cui, W.; Zhang, X.; Wang, Y.; Huang, H.; Chen, B.; Fang, L.; Zhang, H.; Lou,
  J.G.; Zhang, D.
\newblock Text-to-Viz: Automatic Generation of Infographics from
  Proportion-Related Natural Language Statements.
\newblock {\em IEEE Trans. Vis. Comput. Graph.} {\bf
  2019}, {\em 26},~906--916.

\bibitem[Lin \em{et~al.}(2018)Lin, Ford, Adar, and Hecht]{lin2018vizbywiki}
Lin, A.Y.; Ford, J.; Adar, E.; Hecht, B.
\newblock VizByWiki: Mining data visualizations from the web to enrich news
  articles.
\newblock In Proceedings of the 2018 World Wide Web
  Conference, Lyon, France, 23--27 April 2018; pp. 873--882.

\bibitem[Bryan \em{et~al.}(2016)Bryan, Ma, and Woodring]{bryan2016temporal}
Bryan, C.; Ma, K.L.; Woodring, J.
\newblock Temporal summary images: An approach to narrative visualization via
  interactive annotation generation and placement.
\newblock {\em IEEE Trans. Vis. Comput. Graph.} {\bf
  2016}, {\em 23},~511--520.

\bibitem[Tang \em{et~al.}(2017)Tang, Han, Yiu, Ding, and
  Zhang]{tang2017extracting}
Tang, B.; Han, S.; Yiu, M.L.; Ding, R.; Zhang, D.
\newblock Extracting top-k insights from multi-dimensional data.
\newblock In Proceedings of the 2017 ACM International
  Conference on Management of Data, Chicago, IL, USA, 14--19 May 2017; pp. 1509--1524.

\bibitem[Ding \em{et~al.}(2019)Ding, Han, Xu, Zhang, and
  Zhang]{ding2019quickinsights}
Ding, R.; Han, S.; Xu, Y.; Zhang, H.; Zhang, D.
\newblock Quickinsights: Quick and automatic discovery of insights from
  multi-dimensional data.
\newblock In Proceedings of the 2019 International
  Conference on Management of Data, Amsterdam, The Netherlands, 30 June--5 July 2019; pp. 317--332.

\bibitem[Kwon \em{et~al.}(2014)Kwon, Stoffel, J{\"a}ckle, Lee, and
  Keim]{kwon2014visjockey}
Kwon, B.C.; Stoffel, F.; J{\"a}ckle, D.; Lee, B.; Keim, D.
\newblock Visjockey: Enriching data stories through orchestrated interactive
  visualization.
\newblock In Proceedings of the Poster Compendium of the Computation+
  Journalism Symposium, New York, NY, USA, 24--25 October 2014; Volume~3, p.~3.

\bibitem[Luo \em{et~al.}(2018)Luo, Qin, Tang, Li, and Wang]{luo2018deepeye}
Luo, Y.; Qin, X.; Tang, N.; Li, G.; Wang, X.
\newblock Deepeye: Creating good data visualizations by keyword search.
\newblock In Proceedings of the 2018 International
  Conference on Management of Data, Houston, TX, USA, 10--15 June 2018; pp. 1733--1736.

\bibitem[Metoyer \em{et~al.}(2018)Metoyer, Zhi, Janczuk, and
  Scheirer]{metoyer2018coupling}
Metoyer, R.; Zhi, Q.; Janczuk, B.; Scheirer, W.
\newblock Coupling story to visualization: Using textual analysis as a bridge
  between data and interpretation.
\newblock In Proceedings of the 23rd International Conference on Intelligent
  User Interfaces, Tokyo, Japan, 7--11 March 2018; pp. 503--507.

\bibitem[Hoque \em{et~al.}(2017)Hoque, Setlur, Tory, and
  Dykeman]{hoque2017applying}
Hoque, E.; Setlur, V.; Tory, M.; Dykeman, I.
\newblock Applying pragmatics principles for interaction with visual analytics.
\newblock {\em IEEE Trans. Vis. Comput. Graph.} {\bf
  2017}, {\em 24},~309--318.

\bibitem[Kim \em{et~al.}(2020)Kim, Hoque, and Agrawala]{kim2020answering}
Kim, D.H.; Hoque, E.; Agrawala, M.
\newblock Answering questions about charts and generating visual explanations.
\newblock  In Proceedings of the CHI Conference on Human Factors in Computing Systems, Honolulu, HI, USA, 25--30 April 2020; pp. 1--13.

\bibitem[Kafle \em{et~al.}(2020)Kafle, Shrestha, Cohen, Price, and
  Kanan]{kafle2020answering}
Kafle, K.; Shrestha, R.; Cohen, S.; Price, B.; Kanan, C.
\newblock Answering questions about data visualizations using efficient bimodal
  fusion.
\newblock  In Proceedings of the  IEEE/CVF Winter Conference on Applications of Computer Vision, Snowmass Village, CO, USA, 1--5 March 2020; pp. 1498--1507.

\bibitem[Kim \em{et~al.}(2018)Kim, Hoque, Kim, and
  Agrawala]{kim2018facilitating}
Kim, D.H.; Hoque, E.; Kim, J.; Agrawala, M.
\newblock Facilitating document reading by linking text and tables.
\newblock In Proceedings of the 31st Annual ACM Symposium on
  User Interface Software and Technology, Berlin, Germany, 14--17 October 2018; pp. 423--434.

\bibitem[Srinivasan and Stasko(2017)]{srinivasan2017orko}
Srinivasan, A.; Stasko, J.
\newblock Orko: Facilitating multimodal interaction for visual exploration and
  analysis of networks.
\newblock {\em IEEE Trans. Vis. Comput. Graph.} {\bf
  2017}, {\em 24},~511--521.

\bibitem[Mogadala \em{et~al.}(2019)Mogadala, Kalimuthu, and
  Klakow]{mogadala2019trends}
Mogadala, A.; Kalimuthu, M.; Klakow, D.
\newblock Trends in integration of vision and language research: A survey of
  tasks, datasets, and methods.
\newblock {\em arXiv} {\bf 2019},   arXiv:1907.09358

\bibitem[Zhang \em{et~al.}(2009)Zhang, Deng, and Li]{zhang2009concept}
Zhang, J.; Deng, B.; Li, X.
\newblock Concept based query expansion using wordnet.
\newblock In Proceedings of the 2009 International e-Conference on Advanced
  Science and Technology, Daejeon, Korea, 7--9 March 2009; pp. 52--55.

\bibitem[Carpineto and Romano(2012)]{carpineto2012survey}
Carpineto, C.; Romano, G.
\newblock A survey of automatic query expansion in information retrieval.
\newblock {\em ACM Comput. Surv.} {\bf 2012}, {\em 44},~1--50.

\bibitem[Azad and Deepak(2019)]{azad2019query}
Azad, H.K.; Deepak, A.
\newblock Query expansion techniques for information retrieval: A survey.
\newblock {\em Inf. Process. Manag.} {\bf 2019}, {\em
  56},~1698--1735.

\bibitem[Dahir and El~Qadi(2021)]{dahir2021query}
Dahir, S.; El~Qadi, A.
\newblock A query expansion method based on topic modeling and DBpedia
  features.
\newblock {\em Int. J. Inf. Manag. Data Insights}
  {\bf 2021}, {\em 1},~100043.

\bibitem[Hoeber \em{et~al.}(2005)Hoeber, Yang, and
  Yao]{hoeber2005visualization}
Hoeber, O.; Yang, X.D.; Yao, Y.
\newblock Visualization support for interactive query refinement.
\newblock In Proceedings of the  2005 IEEE/WIC/ACM International Conference
  on Web Intelligence (WI'05), Compiegne, France, 19--22 September 2005; pp. 657--665.

\bibitem[Khazaei and Hoeber(2017)]{khazaei2017supporting}
Khazaei, T.; Hoeber, O.
\newblock Supporting academic search tasks through citation visualization and
  exploration.
\newblock {\em Int. J. Digit. Libr.} {\bf 2017}, {\em
  18},~59--72.

\bibitem[Scells and Zuccon(2018)]{scells2018searchrefiner}
Scells, H.; Zuccon, G.
\newblock Searchrefiner: A query visualisation and understanding tool for
  systematic reviews.
\newblock In Proceedings of the 27th ACM International
  Conference on Information and Knowledge Management, Torino, Italy, 22--26 October 2018; pp.~1939--1942.

\bibitem[Russell-Rose and Gooch(2018)]{russell20182dsearch}
Russell-Rose, T.; Gooch, P.
\newblock 2dSearch: A visual approach to search strategy formulation. In Proceedings of the Design of Experimental Search and Information REtrieval Systems (DESIRES 2018), Bertinoro, Italy, 28--31 August 2018.

\bibitem[Curry(2020)]{curry2020dataspaces}
Curry, E., Dataspaces: Fundamentals, {Principles}, and {Techniques}.
\newblock In {\em Real-Time {Linked} {Dataspaces}}; Springer: Berlin/Heidelberg, Germany, 
  2020; pp.
  45--62.

\bibitem[Franklin \em{et~al.}(2005)Franklin, Halevy, and
  Maier]{franklin2005databases}
Franklin, M.; Halevy, A.; Maier, D.
\newblock From databases to dataspaces: a new abstraction for information
  management.
\newblock {\em ACM Sigmod Rec.} {\bf 2005}, {\em 34},~27--33.

\bibitem[Balalau \em{et~al.}(2020)Balalau, Galhardas, Manolescu, Merabti, You,
  Youssef, et~al.]{balalau2020graph}
Balalau, O.; Galhardas, H.; Manolescu, I.; Merabti, T.; You, J.; Youssef, Y.;
  et~al.
\newblock Graph integration of structured, semistructured and unstructured data
  for data journalism.
\newblock {\em arXiv} {\bf 2020},   arXiv:2007.12488

\bibitem[Martinez-Gil(2015)]{martinez2015automated}
Martinez-Gil, J.
\newblock Automated knowledge base management: A survey.
\newblock {\em Comput. Sci. Rev.} {\bf 2015}, {\em 18},~1--9.

\bibitem[Auer \em{et~al.}(2007)Auer, Bizer, Kobilarov, Lehmann, Cyganiak, and
  Ives]{auer2007dbpedia}
Auer, S.; Bizer, C.; Kobilarov, G.; Lehmann, J.; Cyganiak, R.; Ives, Z.
\newblock Dbpedia: A nucleus for a web of open data. In {\em The Semantic Web};
  Springer: Berlin/Heidelberg, Germany, 
  2007; pp. 722--735.

\bibitem[Golshan \em{et~al.}(2017)Golshan, Halevy, Mihaila, and
  Tan]{golshan2017data}
Golshan, B.; Halevy, A.; Mihaila, G.; Tan, W.C.
\newblock Data integration: After the teenage years.
\newblock In Proceedings of the 36th ACM SIGMOD-SIGACT-SIGAI
  Symposium on Principles of Database Systems, Raleigh, NC, USA, 14--19 May 2017; pp. 101--106.

\bibitem[Mountantonakis and Tzitzikas(2019)]{mountantonakis2019large}
Mountantonakis, M.; Tzitzikas, Y.
\newblock Large-scale semantic integration of linked data: A survey.
\newblock {\em ACM Comput. Surv.} {\bf 2019}, {\em 52},~1--40.

\bibitem[Arya \em{et~al.}(2021)Arya, Kuchhal, and Gulati]{arya2021survey}
Arya, A.; Kuchhal, V.; Gulati, K.
\newblock Survey on Data Deduplication Techniques for Securing Data in Cloud
  Computing Environment.
\newblock {\em Smart Sustain. Intell. Syst.} {\bf 2021}, pp.
  443--459.

\bibitem[Christophides \em{et~al.}(2019)Christophides, Efthymiou, Palpanas,
  Papadakis, and Stefanidis]{christophides2019end}
Christophides, V.; Efthymiou, V.; Palpanas, T.; Papadakis, G.; Stefanidis, K.
\newblock End-to-end entity resolution for big data: A survey.
\newblock {\em arXiv} {\bf 2019},  arXiv:1905.06397.

\bibitem[Gröger \em{et~al.}(2014)Gröger, Schwarz, and
  Mitschang]{groger2014deep}
Gröger, C.; Schwarz, H.; Mitschang, B.
\newblock The deep data warehouse: link-based integration and enrichment of
  warehouse data and unstructured content.
\newblock  In Proceedings of the 2014 IEEE 18th International Enterprise Distributed Object Computing Conference, Ulm, Germany , 1--5 September 2014; pp. 210--217.

\bibitem[Roy \em{et~al.}(2005)Roy, Mohania, Bamba, and Raman]{roy2005towards}
Roy, P.; Mohania, M.; Bamba, B.; Raman, S.
\newblock Towards automatic association of relevant unstructured content with
  structured query results.
\newblock  2005, pp. 405--412.

\bibitem[Rosling(accessed in 2022-05-06)]{gapminderusaleviz}
Rosling, H.
\newblock Gapminder---USA's Life Expectancy Line-Chart. Available online: \url{tinyurl.com/gapminderlinechart} (accessed on 6 February 2020). 

\bibitem[Gabbert(accessed in 2022-05-06)]{keyvssq}
Gabbert, E.
\newblock Keywords vs. Search Queries: What’s the Difference?
\newblock
  Available online: \url{https://www.wordstream.com/blog/ws/2011/05/25/keywords-vs-search-queries} (accessed on 6 February 2020).

\bibitem[Everett(2016 (accessed February, 2020))]{elasticsearchwikipedia}
Everett, N.
\newblock Loading Wikipedia's Search Index For Testing.  2016. Available online: \url{https://www.elastic.co/blog/loading-wikipedia} (accessed on 6 February 2020).

\bibitem[Team(2018 (accessed February 15, 2020))]{SimpleElastic}
Team, E.D.
\newblock Simple Query String Query. 
  2018.
\newblock
  Available online: \url{https://www.elastic.co/guide/en/elasticsearch/reference/current/query-dsl-simple-query-string-query.html} (accessed on 15 February 2020).

\bibitem[Feynman(2016)]{feynman2016ebnf}
Feynman, R.
\newblock Ebnf: A Notation to Describe Syntax. 2016.
\newblock {Available online: http://www. ics. uci. edu/\~{}
  pattis/misc/ebnf2. pdf (accessed on 6 May 2022). 
}

\bibitem[Pennington \em{et~al.}(2014)Pennington, Socher, and
  Manning]{pennington2014glove}
Pennington, J.; Socher, R.; Manning, C.D.
\newblock Glove: Global vectors for word representation.
\newblock In Proceedings of the 2014 Conference on Empirical
  Methods in Natural Language Processing (EMNLP), Doha, Qatar, 25--29 October 2014; pp. 1532--1543.

\bibitem[Rehurek and Sojka(2011)]{rehurek2011gensim}
Rehurek, R.; Sojka, P.
\newblock \emph{Gensim--Python Framework for Vector Space Modelling};
\newblock {\em NLP Centre, Faculty of Informatics, Masaryk University: Brno,
  Czech Republic}, {2011}, Volume {3}.

\bibitem[Loper and Bird(2002)]{loper2002nltk}
Loper, E.; Bird, S.
\newblock NLTK: The natural language toolkit.
\newblock {\em arXiv} {\bf 2002},   arXiv:cs/0205028.

\bibitem[Fellbaum(2010)]{fellbaum2010wordnet}
Fellbaum, C.
\newblock WordNet. In {\em Theory and Applications of Ontology: Computer
  Applications}; Springer: Berlin/Heidelberg, Germany, 
  2010; pp. 231--243.

\bibitem[Bhogal \em{et~al.}(2007)Bhogal, MacFarlane, and
  Smith]{bhogal2007review}
Bhogal, J.; MacFarlane, A.; Smith, P.
\newblock A review of ontology based query expansion.
\newblock {\em Inf. Process. Manag.} {\bf 2007}, {\em
  43},~866--886.

\bibitem[Rosling(2018 (accessed February, 2020))]{gapmindergeography}
Rosling, H.
\newblock Geography Related Dataset from Gapminder.  2018.
\newblock Available online: \url{https://www.gapminder.org/data/geo/} (accessed on 6 February 2020).

\bibitem[Brockwell and Davis(2016)]{brockwell2016introduction}
Brockwell, P.J.; Davis, R.A.
\newblock {\em Introduction to Time Series and Forecasting}; Springer: Berlin/Heidelberg, Germany, 
  2016.

\bibitem[Virtanen \em{et~al.}(2020)Virtanen, Gommers, Oliphant, Haberland,
  Reddy, Cournapeau, Burovski, Peterson, Weckesser, Bright,
  et~al.]{virtanen2020scipy}
Virtanen, P.; Gommers, R.; Oliphant, T.E.; Haberland, M.; Reddy, T.;
  Cournapeau, D.; Burovski, E.; Peterson, P.; Weckesser, W.; Bright, J.;
  et~al.
\newblock SciPy 1.0: fundamental algorithms for scientific computing in Python.
\newblock {\em Nat. Methods} {\bf 2020}, {\em 17},~261--272.

\bibitem[Yang \em{et~al.}(2009)Yang, He, and Yu]{Yang2009ComparisonAnalysis}
Yang, C.; He, Z.; Yu, W.
\newblock {Comparison of public peak detection algorithms for MALDI mass
  spectrometry data analysis}.
\newblock {\em BMC Bioinform.} {\bf 2009}, {\em 10},~4.

\bibitem[Lashkari \em{et~al.}(2009)Lashkari, Mahdavi, and
  Ghomi]{lashkari2009boolean}
Lashkari, A.H.; Mahdavi, F.; Ghomi, V.
\newblock A boolean model in information retrieval for search engines.
\newblock In Proceedings of the 2009 International Conference on Information
  Management and Engineering, Kuala Lumpur, Malaysia, 3--5 April 2009; pp. 385--389.

\bibitem[Team(2020 (accessed February, 2020))]{cirrussearch}
Team, W.
\newblock Wikimedia Downloads.  2020. Available online: \url{https://dumps.wikimedia.org/other/cirrussearch/} (accessed on 15 February 2020).


\bibitem[Keogh and Ratanamahatana(2005)]{keogh2005exact}
Keogh, E.; Ratanamahatana, C.A.
\newblock Exact indexing of dynamic time warping.
\newblock {\em Knowl. Inf. Syst.} {\bf 2005}, {\em
  7},~358--386.

\bibitem[M~{\"u} ller(2007)]{muller2007dynamic}
M~{\"u} ller, M.
\newblock Dynamic time warping.
\newblock In {\em Information Retrieval for Music and Motion}; Springer: Berlin/Heidelberg, Germany, 
 {2007}; pp.
  69--84.

\bibitem[Memon \em{et~al.}(2003)Memon, Banerjee, and
  Nagarajan]{Memon2004WhatTesting}
Memon, A.; Banerjee, I.; Nagarajan, A.
\newblock What test oracle should I use for effective GUI testing?
\newblock In Proceedings of the 18th IEEE International Conference on Automated
  Software Engineering, 2003, Montreal, QC, Canada, 6--10 October 2003; pp. 164--173.

\bibitem[Blei \em{et~al.}(2003)Blei, Ng, and Jordan]{blei2003latent}
Blei, D.M.; Ng, A.Y.; Jordan, M.I.
\newblock Latent dirichlet allocation.
\newblock {\em J. Mach. Learn. Res.} {\bf 2003}, {\em
  3},~993--1022.

\bibitem[Reimers and Gurevych(2019)]{reimers-2019-sentence-bert}
Reimers, N.; Gurevych, I.
\newblock Sentence-BERT: Sentence Embeddings using Siamese BERT-Networks.
\newblock In Proceedings of the 2019 Conference on Empirical
  Methods in Natural Language Processing. Association for Computational
  Linguistics, Hong Kong, China, 3--7 November 2019.

\bibitem[Rosling(accessed in 2022-05-06)]{gapminderbubbleviz}
Rosling, H.
\newblock Gapminder---Life Expectancy vs Income Bubble-Chart.
\newblock Available online: \url{https://tinyurl.com/gapminderbubblechart} (accessed on 6 May 2022).


\bibitem[Yu, Jing and Zhang(2020)]{yu2020reasoning}
Yu, Jing; Zhang, Weifeng; Lu, Yuhang; Qin, Zengchang; Hu, Yue; Tan, Jianlong; Wu, Qi.
\newblock Reasoning on the relation: Enhancing visual representation for visual question answering and cross-modal retrieval.
\newblock {\em IEEE Trans. Multimedia.} {\bf
  2020}, {\em 22},~3196--3209.
  
\bibitem[Yu \em{et~al.}(2003)Yu, Zhu, Wang, Zhang, Hu and
  Tan]{yu2020cross}
Yu, Jing; Zhu, Zihao; Wang, Yujing; Zhang, Weifeng; Hu, Yue; Tan, Jianlong.
\newblock Cross-modal knowledge reasoning for knowledge-based visual question answering.
\newblock {\em Pattern Recognition: Elsevier}, {2020}, Volume {108}.

  
\bibitem[Dhelim, Ning and Aung(2020)]{dhelim2020compath}
Dhelim, Sahraoui; Ning, Huansheng; Aung, Nyothiri.
\newblock ComPath: User interest mining in heterogeneous signed social networks for Internet of people.
\newblock {\em IEEE Internet of Things Journal.} {\bf
  2020}, {\em 8},~7024 - 7035.

\end{thebibliography}
\end{document}